\documentclass[prd,preprint,eqsecnum,nofootinbib,amsmath,amssymb,
               tightenlines,dvips,floatfix]{revtex4}
\usepackage{graphicx}
\usepackage{bm}%	bold math
\usepackage{amsfonts}
\usepackage{amssymb}
\newcommand{\mtrx}{\underline}

\usepackage{multirow}
\usepackage{youngtab}
\usepackage{rotating}

\usepackage{dsfont}
\def\openone{\mathds{1}}

\def\b{{\bm b}}

\def\q{{\bm q}}

\def\B{{\bm B}}
\def\calC{{\cal C}}
\def\bcalC{{\bm{{\cal C}}}}
\def\P{{\bm P}}

\def\Nc{N_{\rm c}}

\def\CA{C_{\rm A}}
\def\dA{d_{\rm A}}

\def\tr{\operatorname{tr}}

\def\Tgen{{\mathbb T}}
\def\Adj{{\rm Adj}}
\def\quote#1{``#1"}
\def\quotea#1{``#1"_{\!\!\rm a}}
\def\quotes#1{``#1"_{\!\!\rm s}}
\def\quoteaa#1{``#1"_{\!\!\rm aa}}
\def\quotess#1{``#1"_{\!\!\rm ss}}

\begin {document}

%%%%%%%%%%%%%%%%%%%%%%%%%%%%%%%%%%%%%%%%%%%%%%%%%%%%%%%%%%%%%%%%%%%%%%%%%%%%%%%

%%%%%%%%%%%%%%%%%%%%%%%%%%%%%%%%%%%%%%%%%%%%%%%%%%%%%%%%%%%%%%%%%%%%%%%%%%%%%%%

\title
    {
      The LPM effect in sequential bremsstrahlung:
      from large-\boldmath$N$ QCD to \boldmath$N{=}3$ via
      the SU(\boldmath$N$) analog of Wigner 6-$j$ symbols
    }

\author{Peter Arnold}
%\author{Han-Chih Chang}
\affiliation
    {%
    Department of Physics,
    University of Virginia,
    Charlottesville, Virginia 22904-4714, USA
    \medskip
    }%

\date {\today}

\begin {abstract}%
{%
  Consider a high-energy parton showering as it traverses a QCD medium
  such as a quark-gluon plasma.  Interference effects between
  successive splittings in the shower are potentially very important
  but have so far been calculated (even in idealized theoretical
  situations) only in soft emission or large-$\Nc$ limits, where
  $\Nc$ is the number of quark colors.
  In this paper, we show how one may remove the assumption of
  large $\Nc$ and so begin investigation of $\Nc{=}3$ without
  soft-emission approximations.  Treating finite $\Nc$
  requires (i) classifying
  different ways that four gluons can form a color singlet
  and (ii) calculating medium-induced transitions
  between those singlets,
  for which we find application of results for
  the generalization of Wigner 6-$j$ symbols from angular momentum to
  SU($\Nc$).  Throughout, we make use of the multiple scattering
  ($\hat q$) approximation for high-energy partons crossing quark-gluon
  plasmas, and we find that this approximation is self-consistent
  only if the transverse-momentum diffusion
  parameter $\hat q$ for different color representations satisfies
  Casimir scaling (even for strongly-coupled, and not just weakly-coupled,
  quark-gluon plasmas).
  We also find that results for $\Nc{=}3$
  depend, mathematically, on being able to calculate the
  propagator for a coupled
  non-relativistic quantum harmonic oscillator problem in which the spring
  constants are operators acting on a 5-dimensional Hilbert space of
  internal color states.  Those spring constants are represented by
  constant $5{\times}5$ matrices, which we explicitly construct.
  We are unaware of any closed form solution for this type of
  harmonic oscillator problem, and we discuss
  prospects for using numerical evaluation.
}%
\end {abstract}

\maketitle
\thispagestyle {empty}

%{\def\boldmath{}\tableofcontents}
%\newpage

%%%%%%%%%%%%%%%%%%%%%%%%%%%%%%%%%%%%%%%%%%%%%%%%%%%%%%%%%%%%%%%%%%%%%%%%%%%%%%%

\section{Introduction}
\label{sec:intro}

Very high energy particles traveling through a medium lose
energy primarily through splitting via bremsstrahlung and pair
production.  At very high energy, the quantum mechanical duration of
each splitting process, known as the formation time, exceeds the mean
free time for collisions with the medium, leading to a significant
reduction in the splitting rate known as the Landau-Pomeranchuk-Migdal
(LPM) effect \cite{LP,Migdal}.
In the case of QCD, the basic methods for incorporating
the LPM effect into calculations of splitting rates were developed in
the 1990s by Baier et al.\ \cite{BDMPS12,BDMPS3} and Zakharov
\cite{Zakharov} (BDMPS-Z).
More recently, there has been interest in how to compute important
corrections that arise when two consecutive splittings have
overlapping formation times.%
\footnote{
  For one summary of why this is an interesting problem, see,
  for example, the introduction of ref.\ \cite{qedNfstop}.
}
As we'll
discuss, such calculations must generically address a non-trivial
problem of how to account for the color dynamics of the high-energy
partons as they split while interacting with the medium.  This problem
has been sidestepped in overlapping formation time calculations to
date by taking either (i) the soft bremsstrahlung limit
\cite{Blaizot,Iancu,Wu} or (ii) the large-$\Nc$ limit
\cite{2brem,seq,dimreg,4point}.
In this paper, we show how to avoid these limits by showing how to
incorporate the full color dynamics.  But readers need not
appreciate the full history and formalism of the subject to
follow most of this paper: the problem we most need to
address will be an application of SU($N$) group theory
to (i) the different ways to make color singlets from
four partons and (ii) transitions between those singlets due to
interactions with the medium.  The latter will involve the
SU($N$) generalization of Wigner 6-$j$ coefficients.
The original
purpose of Wigner 6-$j$ coefficients can be thought of as
describing for angular momentum [symmetry group SU(2)] the
relation of different bases for spin singlet states
made from four spins.%
\footnote{
  In textbooks, Wigner 6-$j$ coefficients are
  often described instead as related to the coupling of
  three different angular momenta
  $j_1$, $j_2$, and $j_3$ to
  make a fourth $J$.  (See, for example, section 6.1 of ref.\ \cite{Edmonds}.)
  But since $J$ can be combined with a fourth
  angular momentum $j_4$ to make a singlet if and only if
  $J = j_4$, this is the same problem as how to combine four
  angular momenta $j_1$, $j_2$, $j_3$ and $j_4$ to make a singlet.
}
In our case, we will need the generalization from spin to color.

\begin {figure}[t]
\begin {center}
  \includegraphics[scale=0.6]{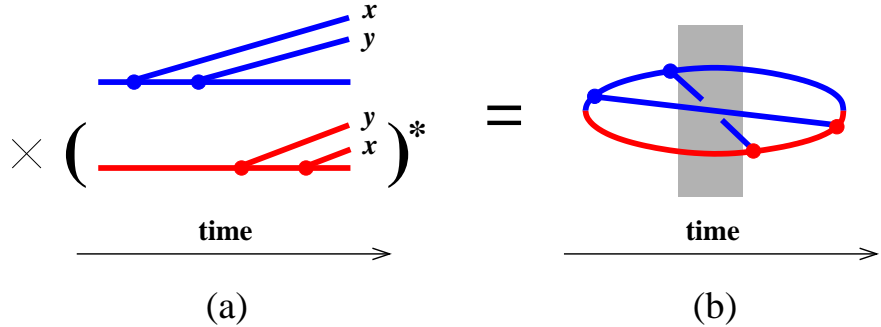}
  \caption{
     \label{fig:split2ab}
     An example of an interference contribution to the rate for
     (overlapping) sequential splitting, with blue indicating the
     amplitude and red the conjugate amplitude.
     In (b), the shading highlights the region of time where
     the evolution must track {\it four} high-energy
     particles (three in the amplitude
     and one in the conjugate amplitude).
  }
\end {center}
\end {figure}

The relevance of studying 4-particle color singlets can be qualitatively
understood by adapting, to the discussion of (overlapping) sequential
splitting, a picture originally used by Zakharov \cite{Zakharov}
to discuss single splitting.  Consider an initial high-energy parton
(quark or gluon) with energy $E$ that splits twice in the medium
to make three high-energy daughters, with energies
$xE$, $yE$ and $(1{-}x{-}y)E$.
Fig.\ \ref{fig:split2ab}a shows
an example of a contribution to the total rate
from an interference
between one way the splitting can happen in the amplitude and
another way it can happen in the conjugate amplitude.  The diagrams
are time-ordered from left to right (and the vertex times should be
integrated over).  Fig.\ \ref{fig:split2ab}b shows
an alternative way of depicting this contribution by sewing the
amplitude and conjugate amplitude diagrams together into a single
diagram for this particular interference contribution to the rate.
Adapting Zakharov's picture, we can recast the calculation by
{\it formally} re-interpreting the right-hand diagram as instead representing
(i) three particles propagating forward in time, followed by a splitting into
(ii) four particles propagating forward in time [shaded region],
followed by a recombination
into (iii) three particle propagating forward in time.%
\footnote{
  There is a technical assumption here that one has integrated the rate over
  the transverse momenta of the final daughters.  See section IV.A of
  ref.\ \cite{2brem} and appendix F of ref.\ \cite{seq} for more details.
}
Only the high-energy particles are explicitly drawn in
fig.\ \ref{fig:split2ab}: each is interacting many times with the medium,
and a statistical average over the medium is performed to get the
(average) rate.  In actual calculations, the evolution of the system
during the three stages (i), (ii), and (iii) just described can be treated
as an effective Schr\"odinger-like problem for the evolution of
the transverse positions of the high-energy particles.
The drawing of fig.\ \ref{fig:split2ab}b suggests, by color conservation
(after medium averaging),
that the high-energy particles should form an
overall color singlet during each stage of evolution.
Because there are multiple ways for four
color charges to form a color singlet, the color dynamics of the
shaded region of fig.\ \ref{fig:split2ab}b can be non-trivial.

In more detail, ref.\ \cite{Vqhat} discusses how the potential energy
for the Schr\"odinger-like evolution equation can be defined in terms of
multi-particle Wilson loops involving lightlike Wilson lines.
These are generalizations of the 2-particle Wilson loops that were
introduced by Liu, Rajagopal, and Wiedemann \cite{LRW1,LRW2} as a
way to define the medium parameters $\hat q_R$ for strongly-interacting
plasmas.  Physically, $\hat q_R$ characterizes the momentum-space
diffusion relation
$\langle Q_\perp^2 \rangle = \hat q_R \, \Delta z$ for the net
transverse momentum kick $\bm{Q}_\perp$ a high-energy particle picks up
from traversing distance $\Delta z$ through the medium, for sufficiently
large $\Delta z$.  Here, $\hat q_R$ depends on the color representation
$R$ of the high-energy particle (e.g.\ fundamental representation for
quarks and adjoint representation for gluons).
In this paper, we start from the result
of ref.\ \cite{Vqhat} that the multi-particle potentials that are
needed for the calculations represented by fig.\ \ref{fig:split2ab}b
are harmonic oscillator potentials that can be written in terms of
$\hat q$ as
\begin {equation}
   \mtrx{V}(\b_1,\b_2,\cdots,\b_n) =
      - \frac{i}8 \sum_{i>j}
         (\hat q_i+\hat q_j-\mtrx{\hat q_{ij}}) (\b_i{-}\b_j)^2
\label {eq:VN}
\end {equation}
in the high energy limit, for which the transverse separations of the
particles are small.  (This high-energy approximation has the same
technical caveats, reviewed in ref.\ \cite{Vqhat}, as most all
other applications of $\hat q$.)
Above, the $\b_i$ are the transverse positions
of the high-energy particles during any particular stage of the evolution.
$\hat q_i$ represents the value of $\hat q$ for the $i$-th particle.
$\hat q_{ij}$ represents the value of $\hat q$ for the color representation
corresponding to the {\it combined}\/ color representation of particles
$i$ and $j$.  In the general case, this combined color representation
is not unique.  For example,
if the $i$ and $j$ represent gluons (which are each in the 8-dimensional
adjoint representation of color), then the combination of the two could
be any irreducible representation in the SU(3) tensor product
$\bm{8}\otimes\bm{8} =
 \bm{1} \oplus \bm{8} \oplus \bm{8}
 \oplus \bm{10} \oplus \overline{\bm{10}} \oplus \bm{27}$.
Since the values of $\hat q$ are mostly different for each
irreducible representation, the $\hat q_{ij}$ in (\ref{eq:VN}) is not
a single number but instead an operator (which can be represented by
a finite-dimensional matrix) acting on the color space of the $N$
particles.  Following ref.\ \cite{Vqhat}, we use underlining
to indicate which quantities are color operators in (\ref{eq:VN}).

Ref.\ \cite{Vqhat} discusses why the color structure of
the $n$-particle potential (\ref{eq:VN}) is trivial for $n{=}3$ because
overall color conservation then implies $\mtrx{\hat q_{12}} = \hat q_3$ and
permutations.  But the color structure is {\it not}\/
trivial for $n{=}4$, which is needed for
evolution of the shaded part of fig.\ \ref{fig:split2ab}b.

One main goal of this paper is to find the explicit
matrix formula for (\ref{eq:VN}) for the case of four gluons, relevant
to the calculation of overlap effects for
double bremsstrahlung $g \to ggg$.  We will review how there are
eight ways to make a color singlet from four gluons.  But we will give
a symmetry argument that three of those singlets decouple from the
color dynamics
of $g \to ggg$, leaving us with a 5-dimensional subspace of
4-gluon color singlet
states, which mix dynamically due to interactions with the medium.
Correspondingly, we will show how $\mtrx{V}(\b_1,\b_2,\b_3,\b_4)$
is explicitly implemented by a $5{\times}5$ matrix function that is
quadratic in transverse separations.  We'll also perform the analysis
for SU($\Nc$) with $\Nc > 3$, to see how
potentials previously used in large-$\Nc$ calculations of overlap effects
are recovered as $\Nc \to \infty$.

With explicit results for the 4-gluon potential in hand,
we demonstrate that consistency of the $\hat q$
approximation for this application necessarily implies that $\hat q$
satisfy Casimir scaling (i.e.\ $\hat q_R \propto C_R$, where $C_R$
is the quadratic Casimir of color representation $R$)
for the specific representations relevant to this paper
(those generated by $\bm{8}\otimes\bm{8}$).
We suspect that Casimir scaling is necessary more generally,
but so far we have not pursued a more general argument.
Though Casimir scaling of $\hat q$ has long been known
through next-to-leading order for weakly-coupled plasmas
\cite{SimonNLO}, we are unaware of a
previous argument for Casimir scaling in applications to
finite-$\Nc$ strongly-coupled plasmas
(again subject to the technical caveats of the $\hat q$ approximation
reviewed in ref.\ \cite{Vqhat}).

The last goal of this paper is to discuss how to use the
explicit potential to calculate
overlap effects for $g\to ggg$ in $\Nc{=}3$ QCD.
Though we will write formulas that can in principle be implemented
numerically, the practical problem of implementing those numerics will be much
more complicated than the case of large-$\Nc$ QCD, and we do
not attempt it today.
We will see that the problem involves finding the propagator
for a quantum system of
two coupled harmonic oscillators whose spring constants are
finite-dimensional
matrices in a space of additional quantum degrees of freedom:
\begin {equation}
   \mtrx{H} =
   \frac{p_1^2}{2 m_1} + \frac{p_2^2}{2 m_2}
   + \frac12
     \begin{pmatrix} q_1 \\ q_2 \end{pmatrix}^{\!\!\top}
     \begin{pmatrix} \mtrx{a} & \mtrx{b} \\ \mtrx{b} & \mtrx{c} \end{pmatrix}
     \begin{pmatrix} q_1 \\ q_2 \end{pmatrix} ,
\label {eq:Hform}
\end {equation}
where here $\mtrx{a}$, $\mtrx{b}$, and $\mtrx{c}$ are
constant ($5{\times}5$) symmetric matrices that do not all commute,
$q_1$ and $q_2$ are two
positions,%
\footnote{
  In our application, the position variables $q_1$ and $q_2$
  will themselves each be two-dimensional vectors $\q_1$ and $\q_2$,
  which we have not made explicit in (\ref{eq:Hform}).
  See (\ref{eq:Hform2}) later for details.
}
$p_1$ and $p_2$ are the corresponding
conjugate momenta, and the ``masses'' $m_1$ and $m_2$ are constants
(unrelated to actual particle masses, which we ignore).
As we'll discuss, the matrix structure of the spring constants makes
computing and using the propagator for the system (\ref{eq:Hform})
more difficult for $\Nc{=}3$ than for the
large-$\Nc$ limit.

% -------------------------------------------------------------------------

\subsection*{Outline}

In the next section, we first discuss the different ways that four gluons can
make an SU(3) color singlet, and the transformations between different basis
choices needed to determine the $\mtrx{\hat q_{ij}}$ in (\ref{eq:VN}) in
terms of the values $\hat q_R$ for different color representations.
Explicit transformations may be found in terms of 6-$j$ coefficients taken
from the literature \cite{NSZ6j,Zakharov6j,Sjodahl}.
We also discuss the permutation symmetries that reduce the
8-dimensional space of singlets to the 5-dimensional subspace relevant for
our application.

In section \ref{sec:Vcasimir}, we find the 4-body potential in the
harmonic approximation (\ref{eq:VN}) by explicitly
constructing the matrices $\mtrx{\hat q_{ij}}$.  We show that consistency
of the color structure of that potential requires that $\hat q_R$ satisfy
Casimir scaling.

In section \ref{sec:reduction}, we briefly review that, for this application,
ref.\ \cite{2brem} showed how to use symmetry to
reduce the 4-body evolution problem to an effective 2-body evolution problem.
We explicitly construct the corresponding harmonic oscillator
Hamiltonian, which we will see has the form (\ref{eq:Hform}).

Section \ref{sec:SUN} discusses the generalization of our analysis to
SU($\Nc$) for $\Nc{>}3$, which has an interesting twist requiring
discussion of how it smoothly matches to $\Nc{=}3$.  We use the
SU($\Nc$) results to see exactly how the color dynamics found
in this paper approaches the simpler large-$\Nc$ results used in
previous work on overlapping formation times.

Section \ref{sec:history} discusses similarities and differences of
this work with earlier work by other authors \cite{NSZ6j,Zakharov6j}.

Section \ref{sec:conclusion} gives our conclusion, where we discuss
the difficulty of matrix-coefficient harmonic oscillator problems
like (\ref{eq:Hform}) and the prospect for numerics.

In the appendix, we outline our own calculation of SU($\Nc$)
6-$j$ coefficients involving four gluons.
%We carried this out
%in part because the list of coefficients we need is not quite
%complete (for $\Nc{>}3$) in the previous literature, and in part
%to independently confirm previous calculations.

% =========================================================================

\section {Color Singlets and 6-\boldmath$j$ coefficients for 4 Gluons}

\subsection {Basics}

\begin {figure}[t]
\begin {center}
  \begin{picture}(305,100)(0,0)
  \put(0,0){\includegraphics[scale=0.5]{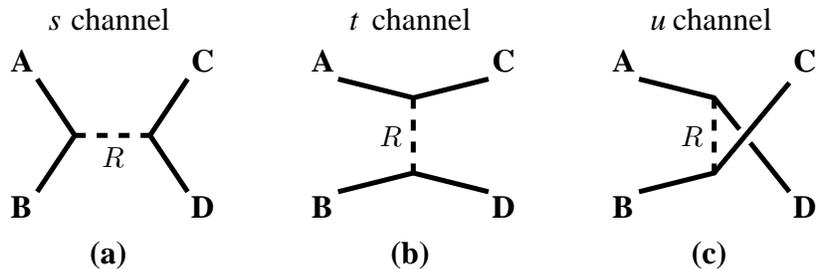}}
  \put(35,40){$R$}
  \put(140,48){$R$}
  \put(254,48){$R$}
  %\put(0,0){.}
  %\put(0,100){.}
  %\put(305,0){.}
  %\put(305,100){.}
  \end{picture}
  \caption{
     \label{fig:channels}
     Different ways of making a basis of singlet states for
     four particles.
  }
\end {center}
\end {figure}

One way to make a basis of color singlet states from 4 particles
(which here we'll call A, B, C, and D) is
depicted in fig.\ \ref{fig:channels}a.
Consider an irreducible color representation $R$ that
appears in both (i) the tensor product $R_{\rm A} \otimes R_{\rm B}$ of
the color representations of particles A and B and (ii) the
tensor product $R_{\rm C} \otimes R_{\rm D}$ of particles C and D.
Combine AB to make $R$ and combine CD to make $R$,
and then combine those two $R$'s to make a singlet.

There is nothing special about choosing to first make the
combinations AB and CD.  We could have just as well chosen
AC and BD, or AD and BC.  We'll refer to these as
$s$-channel, $t$-channel, or $u$-channel choices of bases for
the color singlet states, as depicted in figs.\ \ref{fig:channels}(a--c)
respectively.

\begin {figure}[t]
\begin {center}
  \begin{picture}(355,70)(0,0)
  \put(0,0){\includegraphics[scale=0.5]{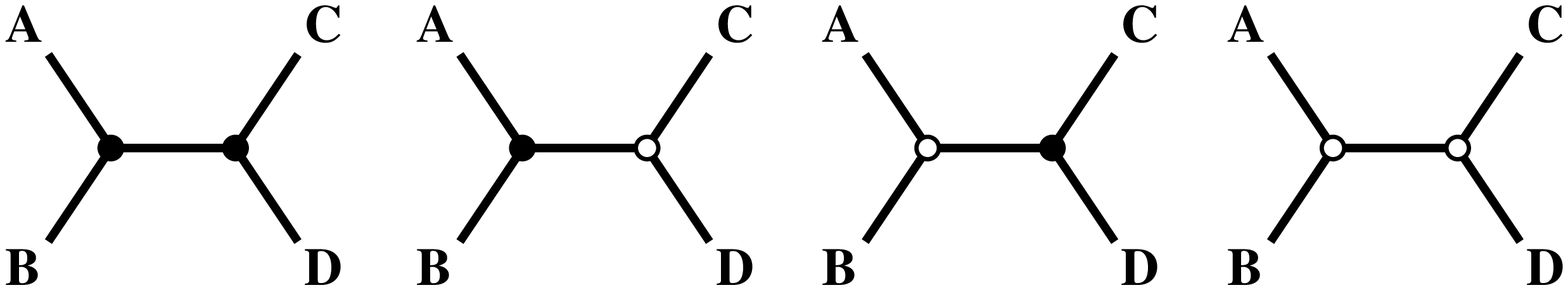}}
  \put(36,36){$\bm{8}$}
  \put(130,36){$\bm{8}$}
  \put(222,36){$\bm{8}$}
  \put(314,36){$\bm{8}$}
  %\put(0,0){.}
  %\put(0,70){.}
  %\put(355,0){.}
  %\put(355,70){.}
  \end{picture}
  \caption{
     \label{fig:schannel8}
     Different ways to make the $R{=}\bm{8}$ case of fig.\ \ref{fig:channels}a.
     Closed and open circles represent the anti-symmetric and symmetric
     combinations $\bm{8}_{\rm a}$ and $\bm{8}_{\rm s}$, respectively.
  }
\end {center}
\end {figure}

The relevant tensor product for combining two gluons is
\begin {equation}
   \bm{8}\otimes\bm{8} =
   \bm{1}_{\rm s} \oplus \bm{8}_{\rm a} \oplus \bm{8}_{\rm s} \oplus
     \bm{10}_{\rm a} \oplus \overline{\bm{10}}_{\rm a} \oplus \bm{27}_{\rm s} ,
\label {eq:8x8}
\end {equation}
where the subscripts give the additional information of whether
each irreducible representation arises in the symmetric (s) or
anti-symmetric (a) combination of $\bm{8}\otimes\bm{8}$.
The decomposition (\ref{eq:8x8}) means that there are five
possibilities
($\bm{1}$, $\bm{8}$, $\bm{10}$, $\overline{\bm{10}}$, and $\bm{27}$)
for the $R$
of fig.\ \ref{fig:channels}a to obtain color singlets from four gluons.
However, there are more than five singlet possibilities because
there are two different ways ($\bm{8}_{\rm a}$ and $\bm{8}_{\rm s}$) each pair
of gluons (AB or CD) can make an $\bm{8}$,
and so there are four different color
singlets corresponding to the case $R{=}\bm{8}$ in fig.
\ref{fig:channels}a.  Following the notation of ref.\ \cite{Sjodahl},
one could denote these four singlets by fig.\ \ref{fig:schannel8},
where a closed circle at a vertex indicates the
anti-symmetric combination $\bm{8}_{\rm a}$ of (\ref{eq:8x8}) and
an open circle represents the symmetric combination $\bm{8}_{\rm s}$.
Because of this multiplicity, there are in total eight independent
color singlets that can be made from four gluons.
The different channels of fig.\ \ref{fig:channels} just
provide different ways to choose a basis for the
eight-dimensional space of 4-gluon singlets.
In what follows, we'll refer to the $s$-channel basis of the
singlet states as
\begin {equation}
   |s_{\bm{1}}\rangle, ~
   |s_{\bm{8}_{\rm aa}}\rangle, ~
   |s_{\bm{8}_{\rm as}}\rangle, ~
   |s_{\bm{8}_{\rm sa}}\rangle, ~
   |s_{\bm{8}_{\rm aa}}\rangle, ~
   |s_{\bm{10}}\rangle, ~
   |s_{\overline{\bm{10}}}\rangle, ~
   |s_{\bm{27}}\rangle ,
\label {eq:sbasis}
\end {equation}
and similarly for the corresponding $t$-channel or $u$-channel bases.

\begin {figure}[ht]
\begin {center}
  \begin{picture}(80,80)(0,0)
  \put(0,0){\includegraphics[scale=0.9]{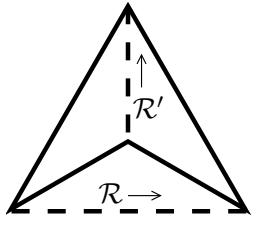}}
  \put(35,5){${\cal R}$}
  \put(48,37){${\cal R'}$}
  %\put(0,0){.}
  %\put(0,80){.}
  %\put(90,0){.}
  %\put(90,80){.}
  \end{picture}
  \caption{
     \label{fig:6j}
     A diagram representing (up to normalization and sign conventions)
     the overlap $\langle s_{\cal R} | t_{{\cal R}'}\rangle$, where solid lines
     represent the adjoint representations for each of our 4 gluons.
     [It also represents the overlap $\langle s_{\cal R} | u_{{\cal R}'} \rangle$,
     depending on how the 4 solid lines are identified with the
     particles ABCD of fig.\ \ref{fig:channels}.]
     When supplemented as necessary by the filled or open circles
     of fig.\ \ref{fig:schannel8} in the case that ${\cal R}$ or
     ${\cal R}'$ is an adjoint representation, these diagrams are
     equivalent to those appearing in ref.\ \cite{Sjodahl}.
     The arrows shown above for ${\cal R}$ and ${\cal R}'$
     are only relevant to the convention for
     distinguishing $\bf{10}$ vs.\ $\overline{\bf{10}}$ or
     $\bm{8}_{as}$ vs.\ $\bm{8}_{sa}$ in this diagram when comparing
     to ref.\ \cite{Sjodahl}.
  }
\end {center}
\end {figure}

The 4-gluon case of the potential (\ref{eq:VN}) is
\begin {align}
   \mtrx{V}(\b_1,\b_2,\b_3,\b_4) =
      - \frac{i}8 \Bigl\{
         &
         (2 \hat q_{\Adj}-\mtrx{\hat q_{12}})
              \bigl[ (\b_1{-}\b_2)^2 + (\b_3{-}\b_4)^2 \bigr]
\nonumber\\ &
         +
         (2 \hat q_{\Adj}-\mtrx{\hat q_{13}})
              \bigl[ (\b_1{-}\b_3)^2 + (\b_2{-}\b_4)^2 \bigr]
\nonumber\\ &
         +
         (2 \hat q_{\Adj}-\mtrx{\hat q_{14}})
              \bigl[ (\b_1{-}\b_4)^2 + (\b_2{-}\b_3)^2 \bigr]
      \Bigr\} ,
\label {eq:V4}
\end {align}
where ``$\Adj$'' means adjoint representation.  Let the particle
labels $1,2,3,4$ here correspond to the $A,B,C,D$ of
fig.\ \ref{fig:channels}.
Then the $s$-channel basis (\ref{eq:sbasis})
diagonalizes $\mtrx{\hat q_{12}}$, since the
$R$ in fig.\ \ref{fig:channels}a is the joint color representation
of the first two particles.
In the $s$-channel basis (\ref{eq:sbasis}),
\begin {equation}
   \mtrx{\hat q_{12}} = \mtrx{\hat q}_{\rm diag} \equiv
   \begin{pmatrix}
      \,0\, &&&&&&& \\
      & \hat q_{\Adj} &&&&&& \\
      && \hat q_{\Adj} &&&&& \\
      &&& \hat q_{\Adj} &&&& \\
      &&&& \hat q_{\Adj} &&& \\
      &&&&& \hat q_{\bm{10}} && \\
      &&&&&& \hat q_{\bm{10}} & \\
      &&&&&&& \hat q_{\bm{27}}
   \end{pmatrix} .
\label {eq:qhat12}
\end {equation}
Similarly, the $t$-channel basis diagonalizes $\mtrx{\hat q_{13}}$, and the
$u$-channel basis diagonalizes $\mtrx{\hat q_{14}}$.  
In order to express the potential (\ref{eq:V4}) explicitly in a single
choice of basis, we need to know how to convert between these
bases choices.  We will choose to express the potential in the
$s$-channel basis, and so we need the unitary matrices
$\mtrx{\langle s|t \rangle}$ and $\mtrx{\langle s|u \rangle}$
of normalized overlaps
\begin {equation}
   \langle s_i | t_j \rangle
   \qquad \mbox{and} \qquad
   \langle s_i | u_j \rangle
\label {eq:overlaps}
\end {equation}
to construct
\begin {equation}
   \mtrx{\hat q_{13}} =
     \mtrx{\langle s|t \rangle} \, \mtrx{\hat q}_{\rm diag} \,
     \mtrx{\langle s|t \rangle}^\dagger ,
   \qquad
   \mtrx{\hat q_{14}} =
     \mtrx{\langle s|u \rangle} \, \mtrx{\hat q}_{\rm diag} \,
     \mtrx{\langle s|u \rangle}^\dagger
\label {eq:qhat13and14}
\end {equation}
in that basis.
Up to normalization and sign conventions, the overlaps
(\ref{eq:overlaps}) are the QCD 6-$j$ coefficients, which can
be represented diagrammatically as in fig.\ \ref{fig:6j},
or as a QCD version
\begin {equation}
  \left\{
    \begin{matrix}
      \Adj & \Adj & {\cal R} \\
      \Adj & \Adj & {\cal R'}
    \end{matrix}
  \right\}
\label {eq:Wigner6j}
\end {equation}
of the Wigner 6-$j$ symbols
\begin {equation}
  \left\{
    \begin{matrix}
      j_1 & j_2 & J \\
      j_3 & j_4 & J'
    \end{matrix}
  \right\}
\end {equation}
for angular momentum.
[In the case where the combined color representation $R$ or $R'$ of a pair
of gluons is an $\bm{8}$, then the ${\cal R}$ or ${\cal R}'$ in
(\ref{eq:Wigner6j}) must also contain the information of whether
that $\bm{8}$ couples as the
$\bm{8}_{\rm aa}$, $\bm{8}_{\rm as}$, $\bm{8}_{\rm sa}$ or $\bm{8}_{\rm ss}$
cases of fig.\ \ref{fig:schannel8}.]

For SU(3), the values of these overlaps can be extracted from
refs.\ \cite{NSZ6j,Zakharov6j} or \cite{Sjodahl}.%
\footnote{
  \label{foot:QCD6j}%
  Ref.\ \cite{NSZ6j} gives the overlaps in their eq.\ (C.17), which
  contains a few misprints that are explicitly corrected for the QCD
  case $\Nc{=}3$ in eq.\ (51) of ref.\ \cite{Zakharov6j}.
  Ref.\ \cite{Sjodahl} gives results for overlaps in their table 3,
  but with a different normalization for states in order to make the
  full tetrahedral symmetry of $6j$-symbols manifest.
  To convert, the entries of ref.\ \cite{Sjodahl} table 3 must
  be multiplied by $[\dim({\cal R}) \, \dim({\cal R}')]^{1/2}$.
  For some relevant earlier work discussing non-abelian $6j$-symbols,
  see refs.\ \cite{Kaplan,Bickerstaff,CvitanovicUn,Cvitanovic}.
}
The values of
$\langle s_i | t_j \rangle$
for conventionally normalized states $|s_i\rangle$ and $|t_j \rangle$,
which gives the unitary transformation between the $t$-channel and
$s$-channel bases for 4-gluon singlets, is shown in table \ref{tab:st3}.
[We have also verified these overlaps independently, which we
briefly summarize in appendix \ref{app:6j} for the sake of
convenient reference.]

\begin{table}[t]

\begin{tabular}{|c|cccccccc|c|}\hline
%\backslashbox{{${\cal R}$}}{{${\cal R}'$}}
$\langle s_{\cal R} | t_{{\cal R}'} \rangle$
  & $\bm{1}$
  & $\bm{8}_{\rm aa}$ & $\bm{8}_{\rm as}$ & $\bm{8}_{\rm sa}$ & $\bm{8}_{\rm ss}$
  & $\bm{10}$ & $\overline{\bm{10}}$
  & $\bm{27}$
  & \shortstack{conversion \\ to $\langle s_{\cal R} | u_{{\cal R}'} \rangle$} \\
\hline
$\bm{1}$
  & $\frac18$
  & $\frac{1}{2\sqrt2}$
  & $0$
  & $0$
  & $\frac{1}{2\sqrt2}$
  & $\frac{\sqrt5}{4\sqrt2}$
  & $\frac{\sqrt5}{4\sqrt2}$
  & $\frac{3\sqrt3}{8}$
  & $+$ \\
$\bm{8}_{\rm aa}$
  & $\frac{1}{2\sqrt2}$
  & $\frac12$
  & $0$
  & $0$
  & $\frac12$
  & $0$
  & $0$
  & $-\frac{\sqrt3}{2\sqrt2}$
  & $-$ \\
$\bm{8}_{\rm as}$
  & $0$
  & $0$
  & $-\frac12$
  & $\frac12$
  & $0$
  & $-\frac12$
  & $\frac12$
  & $0$
  & $+$ \\
$\bm{8}_{\rm sa}$
  & $0$
  & $0$
  & $\frac12$
  & $-\frac12$
  & $0$
  & $-\frac12$
  & $\frac12$
  & $0$
  & $-$ \\
$\bm{8}_{\rm ss}$
  & $\frac{1}{2\sqrt2}$
  & $\frac12$
  & $0$
  & $0$
  & $-\frac{3}{10}$
  & $-\frac{1}{\sqrt5}$
  & $-\frac{1}{\sqrt5}$
  & $\frac{3\sqrt3}{10\sqrt2}$
  & $+$ \\
$\bm{10}$
  & $\frac{\sqrt5}{4\sqrt2}$
  & $0$
  & $-\frac12$
  & $-\frac12$
  & $-\frac{1}{\sqrt5}$
  & $\frac14$
  & $\frac14$
  & $-\frac{\sqrt3}{4\sqrt{10}}$
  & $-$ \\
$\overline{\bm{10}}$
  & $\frac{\sqrt5}{4\sqrt2}$
  & $0$
  & $\frac12$
  & $\frac12$
  & $-\frac{1}{\sqrt5}$
  & $\frac14$
  & $\frac14$
  & $-\frac{\sqrt3}{4\sqrt{10}}$ 
  & $-$ \\
$\bm{27}$
  & $\frac{3\sqrt3}{8}$
  & $-\frac{\sqrt3}{2\sqrt2}$
  & $0$
  & $0$
  & $\frac{3\sqrt3}{10\sqrt2}$
  & $-\frac{\sqrt3}{4\sqrt{10}}$
  & $-\frac{\sqrt3}{4\sqrt{10}}$
  & $\frac{7}{40}$
  & $+$ \\
\hline
\end{tabular}
\caption{
  \label{tab:st3}
  Normalized $st$-channel overlaps
  $\langle s_{\cal R} | t_{{\cal R}'} \rangle$ of
  SU(3) singlets of four adjoint particles, where row labels indicate
  ${\cal R}$ and column labels indicate ${\cal R}'$.
  Conversion to the corresponding $su$-channel overlaps
  $\langle s_{\cal R} | u_{{\cal R}'} \rangle$
  is given by multiplying each row of $\langle s|t \rangle$ by
  the sign shown in the last column above.
}
\end{table}

We should note that changing sign convention for the normalization of
a basis state $|s_{\cal R}\rangle$ would negate the corresponding row
of table \ref{tab:st3}, and a similar change of sign convention for
a $|t_{{\cal R}'}\rangle$ would negate the corresponding column.
So table \ref{tab:st3} corresponds to a specific choice of sign
conventions for the definition of the basis states.

The coefficients $\langle s_{\cal R} | u_{{\cal R'}} \rangle$
giving the transformation from the $u$-basis to the $s$-basis may
now be obtained by swapping the labeling of particles C and D in 
fig.\ \ref{fig:channels} to interchange $t$ channel with $u$ channel.
Specifically, we'll define our $u$ channel singlet
states (including sign convention) by%
\footnote{
  We could have alternatively constructed a definition
  based on A$\leftrightarrow$B, which would differ only by
  sign conventions from (\ref{eq:ubasis}) and in the end would
  produce the same result for the
  $\hat q_{14}$ of (\ref{eq:qhat13and14}).
}
\begin {equation}
   |u_{\cal R} \rangle \equiv
   |t_{\cal R} \rangle ~ \mbox{with C$\leftrightarrow$D} .
\label{eq:ubasis}
\end {equation}
Then
\begin {equation}
  \langle s_{\cal R} | u_{{\cal R}'} \rangle
  =
  \langle s_{\cal R} |
  \Bigl(
   |t_{{\cal R}'} \rangle ~ \mbox{with C$\leftrightarrow$D}
  \Bigr)
  =
  \Bigl(
    \langle s_{\cal R} |  ~ \mbox{with C$\leftrightarrow$D}
  \Bigr)
  |t_{{\cal R}'} \rangle .
\end {equation}
This corresponds to constructing $\mtrx{\langle s|u \rangle}$
from $\mtrx{\langle s|t \rangle}$ by negating those rows
of table \ref{tab:st3} where $C$ and $D$ are combined anti-symmetrically
in $|s_{\cal R}\rangle$.
According to (\ref{eq:8x8}), that's
${\cal R} = \bm{8}_{\rm aa}$, $\bm{8}_{\rm sa}$, $\bm{10}$
and $\overline{\bm{10}}$.
We show the signs corresponding to this conversion in the last
column of table \ref{tab:st3}.

% ------------------------------------------------------------------------

\subsection {Block Diagonalization with Permutation Symmetries}

Because the four representations we are combining into
singlets are identical  (four adjoints),
there are certain permutation symmetries of
ABCD that can be used to simultaneously block-diagonalize
$\mtrx{\langle s| t\rangle}$ and $\mtrx{\langle s| u\rangle}$ and so
reduce the number of singlet states that need be considered.
Specifically, consider the channel-preserving
permutations of ABCD
that simultaneously map $s$-channel to $s$-channel, $t$-channel to $t$-channel,
and $u$-channel to $u$-channel.
From fig.\ \ref{fig:channels}, these are
the symmetries of the rectangle one could draw connecting the labels
ABCD shown in each of those diagrams.
That is, there are two reflection symmetries
\begin {subequations}
\label {eq:rectangle}
\begin {equation}
  \sigma_1: ~ ABCD \leftrightarrow BADC ,
  \qquad
  \sigma_2: ~ ABCD \leftrightarrow CDAB ,
\end {equation}
and the $180^\circ$ rotation symmetry
\begin {equation}
  \sigma_3: ~ ABCD \leftrightarrow DCBA .
\end {equation}
\end {subequations}
The symmetry group of the rectangle (known as $D_2$) is
equivalent to $Z_2 \otimes Z_2$, where any two of the three symmetries
(\ref{eq:rectangle}) may be thought of as the two generators.
In each channel, it will be useful to choose a new basis for our
singlet states, having definite charge
$(\sigma_1,\sigma_2,\sigma_3) = (\pm_1,\pm_2,\pm_3)$
under all three symmetries (\ref{eq:rectangle}).
That's slightly redundant, since
$\sigma_1\sigma_2\sigma_3$ is the identity transformation and so
$\sigma_1\sigma_2\sigma_3 = +1$.
But the redundancy is useful for seeing a simple
relation between charges in different channels:
Consider $ABCD\rightarrow ACBD$, which changes $s$-channel to
$t$-channel, and similarly $ABCD\rightarrow ADBC$ for $s$-channel
to $u$-channel.
Their actions on (\ref{eq:rectangle}) give
\begin {multline}
  \mbox{
    $(\sigma_1,\sigma_2,\sigma_3)=(\sigma_1^s,\sigma_2^s,\sigma_3^s)$
    in $s$-channel construction
  }
\\
  \longrightarrow
  \begin {cases}
    \mbox{
      $(\sigma_1,\sigma_2,\sigma_3)=(\sigma_2^s,\sigma_1^s,\sigma_3^s)$
      in $t$-channel construction;
    } \\
    \mbox{
      $(\sigma_1,\sigma_2,\sigma_3)=(\sigma_2^s,\sigma_3^s,\sigma_1^s)$
      in $u$-channel construction.
    }
  \end {cases}
\end {multline}
For each channel, table \ref{tab:signs} lists ($\sigma_1,\sigma_2,\sigma_3)$
eigenstates and values.

\begin{table}[t]
\begin{tabular}{|c|ccc|}\hline
singlet state
   & \multicolumn{3}{c|}{$(\sigma_1,\sigma_2,\sigma_3)$} \\ \cline{2-4}
(channel ${\rm ch} = s$, $t$, or $u$)
  & $s$-channel & $t$-channel & $u$-channel \\ \hline
$|{\rm ch}_{\bm{1}}\rangle$
   & \multirow{5}{*}{$(+,+,+)$}
   & \multirow{5}{*}{$(+,+,+)$}
   & \multirow{5}{*}{$(+,+,+)$} \\
$|{\rm ch}_{\bm{8}_{\rm aa}}\rangle$
   &&& \\
$|{\rm ch}_{\bm{8}_{\rm ss}}\rangle$
   &&& \\
$\tfrac{1}{\sqrt2} \bigl(
 |{\rm ch}_{\bm{10}}\rangle + |{\rm ch}_{\overline{\bm{10}}}\rangle \bigr)$
   &&& \\
$|{\rm ch}_{\bm{27}}\rangle$
   &&& \\ \hline
$\tfrac{1}{\sqrt2} \bigl(
 |{\rm ch}_{\bm{8}_{\rm as}}\rangle + |{\rm ch}_{\bm{8}_{\rm sa}}\rangle \bigr)$ 
   & $(-,+,-)$ & $(+,-,-)$ & $(+,-,-)$ \\ \hline
$\tfrac{1}{\sqrt2} \bigl(
 |{\rm ch}_{\bm{8}_{\rm as}}\rangle - |{\rm ch}_{\bm{8}_{\rm sa}}\rangle \bigr)$ 
   & $(-,-,+)$ & $(-,-,+)$ & $(-,+,-)$ \\ \hline
$\tfrac{1}{\sqrt2} \bigl(
 |{\rm ch}_{\bm{10}}\rangle - |{\rm ch}_{\overline{\bm{10}}}\rangle \bigr)$
   & $(+,-,-)$ & $(-,+,-)$ & $(-,-,+)$ \\ \hline
\end{tabular}
\caption{
  \label{tab:signs}
  For each channel, the signs $(\sigma_1,\sigma_2,\sigma_3)$ of the
  singlet states under the three channel-preserving permutations
  $(ABCD{\leftrightarrow}BADC, ABCD{\leftrightarrow}CDBA,
    ABCD{\leftrightarrow}DCBA)$.
}
\end{table}

Two states $|s_{\cal R}\rangle$ and $|t_{\cal R'}\rangle$ must have
zero overlap $\langle s_{\cal R}|t_{\cal R'} \rangle$ if their
charges $(\sigma_1,\sigma_2,\sigma_3)$ are different,
and similarly for $|s_{\cal R}\rangle$, $|u_{\cal R'}\rangle$, and
$\langle s_{\cal R}|u_{\cal R'} \rangle$.
That means that the first five rows of table \ref{tab:signs}
do not mix with the last three rows, when one considers
combining interactions mediated by different channels.
Specifically,
the forms (\ref{eq:qhat12}) and (\ref{eq:qhat13and14}) of
$\mtrx{\hat q_{12}}$, $\mtrx{\hat q_{13}}$, and $\mtrx{\hat q_{14}}$
then imply that the 4-particle potential (\ref{eq:V4}) block
diagonalizes into $5{\times}5$ and $3{\times}3$ blocks
corresponding to the first five rows and last
three rows of table \ref{tab:signs}.
We will now see that our application lies within the
$5{\times}5$ block, and so we may ignore the $3{\times}3$ block.

% ------------------------------------------------------------------------

\subsection {Our application is a \boldmath$5{\times}5$ problem}
\label {sec:5x5}

\begin {figure}[t]
\begin {center}
  \includegraphics[scale=0.7]{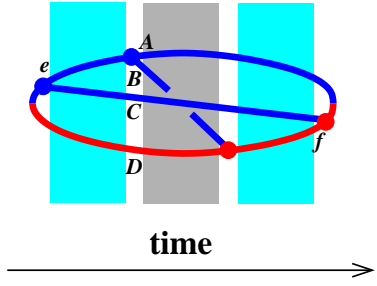}
  \caption{
     \label{fig:split2labels}
     Similar to fig.\ \ref{fig:split2ab} but here showing labels
     for the adjoint color indices of all gluon lines.
  }
\end {center}
\end {figure}

Consider an interference diagram
such as fig.\ \ref{fig:split2ab}b for $g \to ggg$,
which we used to introduce the need for the 4-gluon potential.
We show the diagram again in fig.\ \ref{fig:split2labels}, now
with the adjoint color indices of the gluons labeled, and
with the regions of 3-particle evolution shaded
in light blue (in addition to the 4-particle evolution shaded in
gray).  The important point about this diagram,
true of any interference diagram for $g \to ggg$, is that
3-gluon vertices
are associated with color factors given by
Lie algebra structure constants $f_{abc}$.
In consequence, the three gluons during the first phase of
evolution are in the color singlet state
\begin {equation}
   f_{CDe} | CDe \rangle .
\label {eq:fsinglet}
\end {equation}
None of the interactions with the medium can change this
(after medium averaging),%
\footnote{
  In more detail, there are two possible color singlet states of
  3 gluons: the completely anti-symmetric color combination
  (\ref{eq:fsinglet}) and the completely symmetric one
  $d_{CDe} | CDe \rangle$,
  where $d_{abc}$ is defined in terms of fundamental representation
  generators by $2 \tr(\{T_{\rm F}^a,T_{\rm F}^b\} T_{\rm F}^c)$.
  These two singlets correspond to combining two of the three gluons
  into either the $\bm{8}_{\rm a}$ or $\bm{8}_{\rm s}$ of
  (\ref{eq:8x8}).  In either case, $\mtrx{\hat q_{12}} = \hat q_3$
  for a 3-particle color singlet, and similarly
  $\mtrx{\hat q_{23}} = \hat q_1$ and $\mtrx{\hat q_{31}} = \hat q_2$.
  So the potential
  (\ref{eq:VN}) for 3 gluons is
  $V(\b_1,\b_2,\b_3) =
   -\frac{i}{8} \hat q_{\rm A}
   [(\b_1{-}\b_2)^2 + (\b_2{-}\b_3)^2 + (\b_3{-}\b_1)^2]$,
  which (unlike the 4-body potential) has no interesting color structure
  \cite{Vqhat,2brem}:
  it is the same for both 3-gluon color singlets and so does not induce
  any transitions between the two.
}
and so the color state remains the same just before the second
vertex in fig.\ \ref{fig:split2labels}.
That vertex then splits
$e$ into $AB$ with a color factor of $f_{ABe}$, which means the
intermediate (gray-shaded) region of 4-gluon evolution starts
in color state
\begin {equation}
  f_{ABe} f_{CDe} |ABCD\rangle .
\label {eq:ffstate}
\end {equation}
This corresponds to particles AB being combined anti-symmetrically
into a color adjoint state $f_{ABe} |AB\rangle$ and particles
CD being similarly combined anti-symmetrically into a color
adjoint state $f_{CDe} |CD\rangle$, and then an overall
4-gluon color singlet made from combining the two pairs.
That is, the initial state (\ref{eq:ffstate}) for 4-gluon
evolution is (once normalized) just what we've been calling
$|s_{\bm{8}_{\rm aa}} \rangle$ in this paper.
That places the 4-gluon evolution into the 5-dimensional subspace
given by the first 5 rows of table \ref{tab:signs}.

\begin {figure}[t]
\begin {center}
  \includegraphics[scale=0.7]{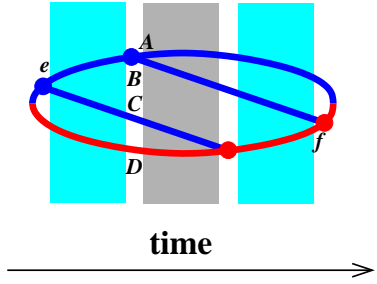}
  \caption{
     \label{fig:seqlabels}
     Similar to fig.\ \ref{fig:split2labels} but for a different
     $g \to ggg$ interference diagram.
  }
\end {center}
\end {figure}

A similar argument working backward from the latest time in
fig.\ \ref{fig:split2labels} shows that the end color state of
the 4-gluon evolution should be taken to be
\begin {equation}
  |t_{\bm{8}_{\rm aa}} \rangle \propto f_{ACe} f_{BDe} |ABCD\rangle .
\end {equation}
Which channel the end 4-gluon state is in compared to the
initial 4-gluon state depends on the interference diagram.
For example, for diagrams like fig.\ \ref{fig:seqlabels},
both will be $|s_{\bm{8}_{\rm aa}} \rangle$.
Regardless, everything will be in the sector of
$(\sigma_1,\sigma_2,\sigma_3){=}(+,+,+)$ singlet states.

The $5{\times}5$ unitary matrix of $\langle s|t \rangle$ overlaps
in the $(\sigma_1,\sigma_2,\sigma_3){=}(+,+,+)$ subspace is shown
explicitly in table \ref{tab:st3block}.

\begin{table}[t]

\begin{tabular}{|c|ccccc|c|}\hline
%\backslashbox{{${\cal R}$}}{{${\cal R}'$}}
$\langle s_{\cal R} | t_{{\cal R}'} \rangle$
  & $\bm{1}$
  & $\bm{8}_{\rm aa}$ & $\bm{8}_{\rm ss}$
  & $\bm{10}{+}\overline{\bm{10}}$
  & $\bm{27}$
  & \shortstack{conversion \\ to $\langle s_{\cal R} | u_{{\cal R}'} \rangle$} \\
\hline
$\bm{1}$
  & $\frac18$
  & $\frac{1}{2\sqrt2}$
  & $\frac{1}{2\sqrt2}$
  & $\frac{\sqrt5}{4}$
  & $\frac{3\sqrt3}{8}$
  & $+$ \\
$\bm{8}_{\rm aa}$
  & $\frac{1}{2\sqrt2}$
  & $\frac12$
  & $\frac12$
  & $0$
  & $-\frac{\sqrt3}{2\sqrt2}$
  & $-$ \\
$\bm{8}_{\rm ss}$
  & $\frac{1}{2\sqrt2}$
  & $\frac12$
  & $-\frac{3}{10}$
  & $-\frac{\sqrt2}{\sqrt5}$
  & $\frac{3\sqrt3}{10\sqrt2}$
  & $+$ \\
$\bm{10}{+}\overline{\bm{10}}$
  & $\frac{\sqrt5}{4}$
  & $0$
  & $-\frac{\sqrt2}{\sqrt5}$
  & $\frac12$
  & $-\frac{\sqrt3}{4\sqrt{5}}$
  & $-$ \\
$\bm{27}$
  & $\frac{3\sqrt3}{8}$
  & $-\frac{\sqrt3}{2\sqrt2}$
  & $\frac{3\sqrt3}{10\sqrt2}$
  & $-\frac{\sqrt3}{4\sqrt{5}}$
  & $\frac{7}{40}$
  & $+$ \\
\hline
\end{tabular}
\caption{
  \label{tab:st3block}
  Normalized $st$-channel overlaps
  $\langle s_{\cal R} | t_{{\cal R}'} \rangle$ for
  the relevant subspace of SU(3) singlets of four adjoint particles.
  Here $\bm{10}{+}\overline{\bm{10}}$ is shorthand for the
  $\tfrac{1}{\sqrt2} \bigl(
   |{\rm ch}_{\bm{10}}\rangle + |{\rm ch}_{\overline{\bm{10}}}\rangle \bigr)$
  state of table \ref{tab:signs}.
  The last column shows how to convert to
  $\langle s_{\cal R} | u_{{\cal R}'} \rangle$ just as in
  table \ref{tab:st3}.
}
\end{table}

% =========================================================================

\section {The explicit potential and Casimir scaling}
\label {sec:Vcasimir}

\subsection {The 4-body potential}

Restricting attention to the $5$-dimensional subspace of relevant
singlets, the 4-gluon potential is now given by (\ref{eq:V4})
with
\begin {subequations}
\label {eq:V4bits}
\begin {equation}
   \mtrx{\hat q_{12}} = \mtrx{\hat q}_{\rm diag} \equiv
   \begin{pmatrix}
      \,0\, &&&& \\
      & \hat q_{\Adj} &&& \\
      && \hat q_{\Adj} && \\
      &&& \hat q_{\bm{10}} & \\
      &&&& \hat q_{\bm{27}}
   \end{pmatrix} ,
\end {equation}
\begin {equation}
   \mtrx{\hat q_{13}} =
     \mtrx{\langle s|t \rangle} \, \mtrx{\hat q}_{\rm diag} \,
     \mtrx{\langle s|t \rangle}^\dagger ,
\end {equation}
and
\begin {equation}
   \mtrx{\hat q_{14}} =
     \mtrx{\langle s|u \rangle} \, \mtrx{\hat q}_{\rm diag} \,
     \mtrx{\langle s|u \rangle}^\dagger
   =
   \mtrx{P} \, \mtrx{\hat q_{13}} \, \mtrx{P} ,
\end {equation}
where $\mtrx{\langle s|t \rangle}$ is given by table
\ref{tab:st3block}, and
$\mtrx{P}$ is a diagonal matrix composed of the signs
in the last column of that table:
\begin {equation}
  \mtrx{P} \equiv
  \begin{pmatrix}
      $+$ &&&& \\
      & $-$ &&& \\
      && $+$ && \\
      &&& $-$ & \\
      &&&& $+$
   \end{pmatrix} .
\end {equation}
\end {subequations}

% ---------------------------------------------------------------------------

\subsection {An additional consistency condition}

Ref.\ \cite{Vqhat} showed that consistency of the harmonic oscillator
($\hat q$) approximation to the $n$-body potential requires that
approximation to have the form (\ref{eq:VN}) and so, in the 4-body
case, requires (\ref{eq:V4}).  However, ref.\ \cite{Vqhat} did
not show the inverse: it did not prove that (\ref{eq:VN})
is always consistent.  ``Consistency'' here means
that the $n$-body potential matches the $(n{-}1)$-body potential
in the special cases where two of the particles
are in the same place (i.e.\ $\b_i{=}\b_j$ for some $i$ and $j$).
We're now in a position to check this for the 4-gluon potential
given by (\ref{eq:V4}), (\ref{eq:V4bits}), and table \ref{tab:st3block}.
Consider the special case where $\b_3{=}\b_4$ in (\ref{eq:V4}),
giving
\begin {equation}
   \mtrx{V}(\b_1,\b_2,\b_3,\b_3) =
   - \frac{i}8 \Bigl\{
         (2 \hat q_{\Adj}-\mtrx{\hat q_{12}}) (\b_1{-}\b_2)^2
         +
         (4 \hat q_{\Adj}-\mtrx{\hat q_{13}}-\mtrx{\hat q_{14}})
              \bigl[ (\b_1{-}\b_3)^2 + (\b_2{-}\b_3)^2 \bigr]
   \Bigr\} .
\label {eq:V4special}
\end {equation}
However, as in ref.\ \cite{Vqhat}, we may alternatively think of
this situation
as a {\it three}-particle problem
by replacing coincident particles 3 and 4 above by
a single particle with their combined color representation, i.e.\
by replacing $\hat q_3 \to \mtrx{\hat q_{34}}$ in the formula
for a 3-body particles potential.

Generically, the 3-body case of the $n$-body potential (\ref{eq:VN})
is \cite{Vqhat,2brem}
\begin {equation}
   \mtrx{V}(\b_1,\b_2,\b_3) =
   - \frac{i}8 \Bigl\{
         (\hat q_1 + \hat q_2 - \hat q_3) (\b_1{-}\b_2)^2
         +
         (\hat q_2 + \hat q_3 - \hat q_1) (\b_2{-}\b_3)^2
         +
         (\hat q_3 + \hat q_1 - \hat q_2) (\b_3{-}\b_1)^2
   \Bigr\} ,
\label {eq:V3}
\end {equation}
where the fact that the potential is for an overall 3-body
color singlet has been used to replace the $\mtrx{\hat q_{12}}$ of
(\ref{eq:VN}) by $\hat q_3$ in (\ref{eq:V3}), with similar
replacements for
$\mtrx{\hat q_{13}}$ and $\mtrx{\hat q_{23}}$.
Consistency of (\ref{eq:V4special}) and (\ref{eq:V3})
then implies that we must have
\begin {equation}
   \mtrx{V}(\b_1,\b_2,\b_3,\b_3) =
   \Bigl[
     V(\b_1,\b_2,\b_3)
     ~\mbox{with $\hat q_3 \to \mtrx{\hat q_{34}}$}
   \Bigr] . 
\label {eq:Vconsistency}
\end {equation}
Ref.\ \cite{Vqhat} focused on the consistency of what, in
the case of (\ref{eq:V4special}), would be the $(\b_1{-}\b_2)^2$
term.  Appropriately generalized, that comparison
was the basis for how the
$n$-body potential (\ref{eq:VN}) was determined.
Here, let's also check the consistency of the {\it other} terms in
(\ref{eq:V4special}).
Comparing (\ref{eq:V4special}) to (\ref{eq:V3}) via
(\ref{eq:Vconsistency}), we see that
full consistency of the 4-gluon potential is achieved if and only if
additionally
\begin {equation}
  4 \hat q_\Adj-\mtrx{\hat q_{13}}-\mtrx{\hat q_{14}} = \mtrx{\hat q_{34}} .
\label {eq:consistency0}
\end {equation}
$\mtrx{\hat q_{34}} = \mtrx{\hat q_{12}}$ since the 4 gluons
form a color singlet, and so we can rewrite the consistency
condition (\ref{eq:consistency0}) as
\begin {equation}
  \mtrx{\hat q_{12}} + \mtrx{\hat q_{13}} + \mtrx{\hat q_{14}}
  = 4\hat q_\Adj .
\label {eq:consistency}
\end {equation}

This same condition could alternatively be derived by looking at
the special case $\b_2{=}\b_3{=}\b_4$ of the 4-gluon potential
(\ref{eq:V4}),
\begin {equation}
   \mtrx{V}(\b_1,\b_2,\b_2,\b_2) =
   - \frac{i}8 
         (6 \hat q_{\Adj}-\mtrx{\hat q_{12}}-\mtrx{\hat q_{13}}-\mtrx{\hat q_{14}})
              (\b_1{-}\b_2)^2
   \Bigr\} .
\label {eq:V4veryspecial}
\end {equation}
By overall color neutrality, the combination of the coincident particles
2, 3, and 4 must be in the adjoint representation, and so
(\ref{eq:V4veryspecial}) must match up with the 2-gluon potential
$V(\b_1,\b_2) = -\frac{i}8 (2\hat q_\Adj) (\b_1-\b_2)^2$.
That matching also gives (\ref{eq:consistency}).

% ---------------------------------------------------------------------------

\subsection{Consistency implies Casimir Scaling of \boldmath$\hat q$}

Using the explicit formulas of (\ref{eq:V4bits}) and
table \ref{tab:st3block}, we find that  (\ref{eq:consistency})
is satisfied only if the $\hat q_R$ for the three non-trivial
representations $R=\bm{8}$, $\bm{10}$, and $\bm{27}$ appearing
in our analysis are related by Casimir scaling.  Namely,
consistency requires
\begin {equation}
  \hat q_{\bm{10}} = \frac{C_{\bm{10}}}{C_\Adj} \, \hat q_\Adj
    = 2 \hat q_\Adj
  \quad \mbox{and} \quad
  \hat q_{\bm{27}} = \frac{C_{\bm{27}}}{C_\Adj} \, \hat q_\Adj
    = \tfrac{8}{3} \hat q_\Adj .
\end {equation}
We speculate that there is a general argument that applies to
other representations as well, but in this paper we focus
only on the application to 4-gluon potentials.

Casimir scaling implies that
\begin {equation}
  \mtrx{\hat q_{ij}}
  = \frac{\mtrx{C_{ij}}}{C_\Adj} \, \hat q_\Adj
  = \frac{(\mtrx{\Tgen}_i + \mtrx{\Tgen}_j)^2}{C_\Adj} \, \hat q_\Adj
  = \frac{C_i + 2\mtrx{\Tgen}_i\cdot\mtrx{\Tgen}_j + C_j}{C_\Adj} \, \hat q_\Adj ,
\end {equation}
where (following the notation of ref.\ \cite{Vqhat})
$\mtrx{\Tgen}_i^a$ are the color generators acting on particle $i$, and
$\mtrx{\Tgen}_i\cdot\mtrx{\Tgen}_j \equiv \mtrx{\Tgen}^a_i\cdot\mtrx{\Tgen}^a_j$.
The 4-gluon potential (\ref{eq:V4}) can then be rewritten as
\begin {align}
   \mtrx{V}(\b_1,\b_2,\b_3,\b_4) =
      \frac{i\hat q_\Adj}{4} \biggl\{
         &
         \frac{\mtrx{\Tgen}_1\cdot\mtrx{\Tgen}_2}{C_\Adj}
              \bigl[ (\b_1{-}\b_2)^2 + (\b_3{-}\b_4)^2 \bigr]
\nonumber\\ &
         +
         \frac{\mtrx{\Tgen}_1\cdot\mtrx{\Tgen}_3}{C_\Adj}
              \bigl[ (\b_1{-}\b_3)^2 + (\b_2{-}\b_4)^2 \bigr]
\nonumber\\ &
         +
         \frac{\mtrx{\Tgen}_1\cdot\mtrx{\Tgen}_4}{C_\Adj}
              \bigl[ (\b_1{-}\b_4)^2 + (\b_2{-}\b_3)^2 \bigr]
      \biggr\} ,
\label {eq:V4Casimir}
\end {align}
which has the same form as the weakly-coupled plasma result discussed
in \cite{Vqhat} but here is now a feature for $\hat q$ approximations
in strongly-coupled plasmas as well.

We can relate the ${\Tgen}_i\cdot{\Tgen}_j$ of (\ref{eq:V4Casimir})
back to (\ref{eq:V4bits}) for the $\mtrx{\hat q_{ij}}$
via
\begin {equation}
  \frac{\mtrx{\Tgen}_i\cdot\mtrx{\Tgen}_j}{C_\Adj}
  = \frac{\mtrx{\hat q_{ij}}}{2 \hat q_\Adj} - \mtrx{\openone} .
\end {equation}
This yields explicit results%
\footnote{
  Our potential of (\ref{eq:V4Casimir}--\ref{eq:ST}) is closely related
  to that of ref.\ \cite{NSZ6j} eqs.\ (59) and (70).  See our
  section \ref{sec:history} for more discussion.
}
\begin {subequations}
\label {eq:TdotTST}
\begin {equation}
   \frac{\mtrx{\Tgen}_1\cdot\mtrx{\Tgen}_2}{C_\Adj}
     = -\mtrx{S} \,,
   \qquad
   \frac{\mtrx{\Tgen}_1\cdot\mtrx{\Tgen}_3}{C_\Adj}
     = \tfrac12 (\mtrx{S}-\mtrx{\openone}) - \mtrx{T} \,,
   \qquad
   \frac{\mtrx{\Tgen}_1\cdot\mtrx{\Tgen}_4}{C_\Adj}
     = \tfrac12 (\mtrx{S}-\mtrx{\openone}) + \mtrx{T} \,,
\label {eq:TdotT}
\end {equation}
where
\begin {equation}
   \mtrx{S} \equiv
     \begin{pmatrix}
       ~1 &&&& \\
       & ~\tfrac12 &&& \\
       && ~\tfrac12 && \\
       &&& ~0 & \\
       &&&& -\tfrac13
     \end{pmatrix} ,
   \qquad
   \mtrx{T} \equiv
     \begin{pmatrix}
       0 & \tfrac{1}{2\sqrt2} & 0 & 0 & 0 \\
       \frac{1}{2\sqrt2} & 0 & \tfrac14 & 0 & \tfrac{1}{2\sqrt6} \\
       0 & \tfrac14 & 0 & \tfrac{1}{\sqrt{10}} & 0 \\
       0 & 0 & \tfrac{1}{\sqrt{10}} & 0 & \tfrac{1}{\sqrt{15}} \\
       0 & \tfrac{1}{2\sqrt6} & 0 & \tfrac{1}{\sqrt{15}} & 0
     \end{pmatrix} .
\label {eq:ST}
\end {equation}
\end {subequations}

% ===========================================================================

\section{Reduction of 4-body to effective 2-body problem}
\label{sec:reduction}

For applications to overlapping formation times,
ref.\ \cite{2brem} showed that symmetries and conservation laws could
be used to mathematically reduce the 4-body evolution problem to
an effective 2-body evolution problem.%
\footnote{
  See in particular section III of ref.\ \cite{2brem}.
}
In the notation of ref.\ \cite{2brem}, we can rewrite differences of
$(\b_1,\b_2,\b_3,\b_4)$ in terms of $(\bcalC_{12},\bcalC_{34})$ with%
\footnote{
  Our $\bcalC_{ij}$ here is called ${\bm C}_{ij}$ in ref.\ \cite{2brem}.
  Here we use the notation $\bcalC_{ij}$ to help avoid
  confusion with the Casimir operator $\mtrx{C_{ij}}$ for the joint color
  representation of particles $i$ and $j$.
}
\begin {equation}
   \bcalC_{ij} \equiv \frac{\b_i{-}\b_j}{x_i{+}x_j} ,
\end {equation}
where $x_i$ is the longitudinal momentum fraction of particle
$i$ in figs.\ \ref{fig:split2labels} or \ref{fig:seqlabels},
relative to the longitudinal momentum of the initial gluon that
started the $g \to ggg$ double splitting process.
The $x_i$ are defined
with a sign convention for particles in the conjugate
amplitude so that $x_1 + x_2 + x_3 + x_4 = 0$.
The relations are%
\footnote{
  Eqs.\ (5.14) of ref.\ \cite{2brem}.
}
\begin {subequations}
\label {eq:brelations}
\begin {align}
   \b_1 - \b_2 &= (x_1+x_2) \bcalC_{12} ,
\\
   \b_1 - \b_3 &= x_2 \bcalC_{12} - x_4 \bcalC_{34} ,
\\
   \b_1 - \b_4 &= x_2 \bcalC_{12} + x_3 \bcalC_{34} ,
\\
   \b_2 - \b_3 &= - x_1 \bcalC_{12} - x_4 \bcalC_{34} ,
\\
   \b_2 - \b_4 &= - x_1 \bcalC_{12} + x_3 \bcalC_{34} , 
\\
   \b_3 - \b_4 &= (x_3+x_4) \bcalC_{34} .
\end {align}
\end {subequations}
Eqs.\ (\ref{eq:V4Casimir}) and (\ref{eq:TdotT}) then give
\begin {subequations}
\label {eq:HVeffective}
\begin {multline}
  \mtrx{V}(\bcalC_{12},\bcalC_{34})
  =
  - \tfrac{i}{4} \hat q_\Adj
  \Bigl\{
     (x_1^2 + 2 x_1 x_2 \mtrx{S} + x_2^2) \calC_{12}^2
     +
     (x_3^2 + 2 x_3 x_4 \mtrx{S} + x_4^2) \calC_{34}^2
\\
     +
     2
     \bigl[
        \tfrac12 (x_1-x_2) (x_3-x_4) (\mtrx{S}-1)
        - (x_1+x_2)(x_3+x_4) \mtrx{T}
     \bigr] \bcalC_{12}\cdot\bcalC_{34}
   \Bigr\} .
\label {eq:Veffective}
\end {multline}
Ref.\ \cite{2brem} shows that the Schr\"odinger-like problem
describing 4-particle evolution has Hamiltonian%
\footnote{
  Specifically, (\ref{eq:Heffective}) here
  is the Hamiltonian corresponding to the
  Lagrangian displayed in
  eqs.\ (5.15--16) of ref.\ \cite{2brem}.
}
\begin {equation}
  \mtrx{H} =
  \frac{P_{12}^2}{2 x_1 x_2 (x_1{+}x_2) E}
  +
  \frac{P_{34}^2}{2 x_3 x_4 (x_3{+}x_4) E}
  +
  \mtrx{V}(\bcalC_{12},\bcalC_{34}) ,
\label {eq:Heffective}
\end {equation}
\end {subequations}
except that the potential $V(\bcalC_{12},\bcalC_{34})$ there has been
replaced by a $5{\times}5$ matrix-valued potential here,
as indicated by the underlining above.
$\P_{12}$ and $\P_{34}$ are the momenta conjugate to
$\bcalC_{12}$ and $\bcalC_{34}$, and $E$ is the energy of the initial gluon
in the $g \to ggg$ double splitting process.

The Hamiltonian (\ref{eq:Heffective}) is of the form (\ref{eq:Hform})
previewed in the introduction except that the position variables
$q_1$ and $q_2$ are two-dimensional transverse position vectors
$\bcalC_{12}$ and $\bcalC_{34}$.  So (\ref{eq:Hform}) is more explicitly
of the form
\begin {equation}
   \mtrx{H} =
   \sum_{i=x,y}
   \left[
     \frac{p_{1i}^2}{2 m_1} + \frac{p_{2i}^2}{2 m_2}
     + \frac12
       \begin{pmatrix} q_{1i} \\ q_{2i} \end{pmatrix}^{\!\!\top}
       \begin{pmatrix} \mtrx{a} & \mtrx{b} \\ \mtrx{b} & \mtrx{c} \end{pmatrix}
       \begin{pmatrix} q_{1i} \\ q_{2i} \end{pmatrix}
   \right] ,
\label {eq:Hform2}
\end {equation}
where $i$ runs over the two transverse dimensions.  In our application,
\begin {align}
  m_1 &= x_1 x_2 (x_1+x_2) E ,
\\
  m_2 &= x_3 x_4 (x_3+x_4) E ,
\end {align}
and
\begin {align}
  \mtrx{a} &= -\tfrac{i}{2} \hat q_\Adj (x_1^2 + 2 x_1 x_2 \mtrx{S} + x_2^2) ,
\\
  \mtrx{b} &= -\tfrac{i}{2} \hat q_\Adj
     \bigl[
        \tfrac12 (x_1-x_2) (x_3-x_4) (\mtrx{S}-1)
        - (x_1+x_2)(x_3+x_4) \mtrx{T}
     \bigr] ,
\\
  \mtrx{c} &= -\tfrac{i}{2} \hat q_\Adj (x_3^2 + 2 x_3 x_4 \mtrx{S} + x_4^2) .
\end {align}

If it were not for the
color structure (the matrix nature of the spring constants $\mtrx{a}$,
$\mtrx{b}$, $\mtrx{c}$), the evolution of the $x$ components
$(q_{1x},q_{2x})$ would be independent of the evolution of the $y$
components $(q_{1y},q_{2y})$, and so it would be enough to solve the
problem for a coupled pair of {\it one}-dimensional oscillators.
However, in (\ref{eq:Hform2}), the $(q_{1x},q_{2x})$ part of the Hamiltonian
could change the color state of the system, which would then in turn
affect the evolution of the $(q_{1y},q_{2y})$ degrees of freedom, and
vice versa.  So the evolution of $(q_{1x},q_{2x})$ is not independent
from that of $(q_{1y},q_{2y})$.

The calculations \cite{2brem} of overlapping splitting make use of
the propagator $G(\bcalC'_{12},\bcalC'_{34},t;\bcalC_{12},\bcalC_{34},0)$
for 4-particle evolution.  Recall from section
\ref{sec:5x5} that the 4-particle evolution starts with the
singlet state $|{\rm ch}_{\bm{8}_{\rm aa}}\rangle$ in some channel
(${\rm ch} = s$, $t$, or $u$, depending on how we label the particles)
and similarly ends with the singlet state $|{\rm ch}'_{\bm{8}_{\rm aa}}\rangle$
in some channel (${\rm ch}' = s$, $t$, or $u$, depending on the
interference diagram).  The relevant propagator matrix elements for the
application to overlapping formation times are then
\begin {equation}
   \langle{\rm ch}'_{\bm{8}_{\rm aa}}|
   \, \mtrx{G}(\bcalC'_{12},\bcalC'_{34},t;\bcalC_{12},\bcalC_{34},0) \,
   |{\rm ch}_{\bm{8}_{\rm aa}}\rangle .
\label {eq:propagator}
\end {equation}
In the $s$-channel basis
$\bigl(
  |s_{\bm{1}}\rangle, |s_{\bm{8}_{\rm aa}}\rangle, |s_{\bm{8}_{\rm ss}}\rangle,
  |s_{\bm{10}+\overline{\bm{10}}}\rangle, |s_{\bm{27}}\rangle
\bigr)$
used for our matrices
in (\ref{eq:ST}),
\begin {equation}
  |s_{\bm{8}_{\rm aa}}\rangle \to
    \begin{pmatrix} 0 \\ 1 \\ 0 \\ 0 \\ 0 \end{pmatrix}
  \qquad
  \mbox{and}
  \qquad
  |t_{\bm{8}_{\rm aa}}\rangle \to
    \mtrx{\langle s|t \rangle}
    \begin{pmatrix} 0 \\ 1 \\ 0 \\ 0 \\ 0 \end{pmatrix}
  =
  \begin{pmatrix}
    \tfrac{1}{2\sqrt2} \\ \tfrac12 \\ \tfrac12 \\ 0 \\ -\tfrac{\sqrt3}{2\sqrt2}
  \end{pmatrix} .
\label {eq:stvecs}
\end {equation}
We will explain in the conclusion why the color matrix structure
of the propagator $\mtrx{G}$ creates
a much more difficult problem for numerical calculations of
overlapping splittings than for the large-$\Nc$ case \cite{2brem}.

% ===========================================================================

\section{SU(\boldmath$N$) and recovering the large-\boldmath$N$ limit}
\label{sec:SUN}

We will now generalize the preceding results from SU(3) to SU($N$), with
the goal of studying how the (simpler) large-$N$ formalism for
earlier calculations \cite{2brem,seq,dimreg,4point}
of overlapping formation time effects is recovered as $N{\to}\infty$.
From here on, we refer to the number of colors $\Nc$ as simply $N$
to make formulas more compact and easy to read.

% ----------------------------------------------------------------------------

\subsection{SU(\boldmath$N$) results}

An interesting feature of SU($N$) for $N > 3$ is that one more
irreducible representation appears in the tensor product $\Adj\otimes\Adj$
than for SU(3).  Reassuringly, we
will find that the limit $N\to 3$ of the potential nonetheless smoothly
approaches the SU(3) case.  In terms of Young Tableaux, the tensor
product $\Adj\otimes\Adj$ is
\begin{align}
\Yvcentermath1
\Ystdtext1
 \mbox{\begin{sideways}$\kern-1em N{-}1$\end{sideways}}
    \left\updownarrow ~ \yng(2,1,1,1,1,1) \right.
 \otimes \yng(2,1,1,1,1,1) =
 1 \oplus 
 \yng(2,1,1,1,1,1) \oplus 
 \yng(2,1,1,1,1,1) \oplus
 ~
 \mbox{\begin{sideways}$\kern-1em N{-}2$\end{sideways}}
    \left\updownarrow ~ \yng(3,1,1,1,1) \right. \oplus
 \yng(3,3,2,2,2,2) \oplus
 \yng(4,2,2,2,2,2) \oplus
 \yng(2,2,1,1,1)
\label {eq:tableux}
\end {align}
with corresponding dimensions
\begin {multline}
  (N^2{-}1)\otimes(N^2{-}1) =
\\
  1_s \oplus
  (N^2{-}1)_{\rm a} \oplus
  (N^2{-}1)_{\rm s} \oplus
  \left(\frac{(N^2{-}4)(N^2{-}1)}{4}\right)_{\rm a} \oplus
  \left(\overline{\frac{(N^2{-}4)(N^2{-}1)}{4}}\right)_{\rm a}
\\
  \oplus
  \left(\frac{N^2(N{-}1)(N{+}3)}{4}\right)_{\rm s} \oplus
  \left(\frac{N^2(N{+}1)(N{-}3)}{4}\right)_{\rm s} .
\label {eq:SUNdims}
\end {multline}
For the sake of easy comparison to previous SU(3) results,
we'll refer to these representations by their $N{=}3$ dimensions
but use scare quotes as a reminder that we're really considering
$N > 3$:
\begin {equation}
 \quote{\bm{8}} \otimes \quote{\bm{8}} =
 \bm{1}_{\rm s} \oplus \quotea{\bm{8}} \oplus \quotes{\bm{8}}
 \oplus \quotea{\bm{10}} \oplus \quotea{\overline{\bm{10}}}
 \oplus \quotes{\bm{27}}
 \oplus \quotes{\bm{0}} \,.
\label {eq:AdjxAdj}
\end {equation}
The extra irreducible representation for $N{>}3$ is the one labeled
$\quote{\bm{0}}$ above, and the corresponding Young Tableaux in
(\ref{eq:tableux}) is illegal for SU(3).
However, a first point of reassurance concerning using $N{>}3$ results
to study the transition from $N{\to}\infty$ to $N{=}3$ is that
the dimension of this unwanted
representation shrinks to zero as $N{\to}3$.

In appendix \ref{app:6j}, we review explicit constructions of (i) the
irreducible representations in
the decomposition (\ref{eq:AdjxAdj}) and (ii) the corresponding 4-gluon
singlet states.  Similar to the $|{\rm ch}_{\bm{27}}\rangle$ of
table \ref{tab:signs}, singlets $|{\rm ch}_{\quote{\bm{0}}}\rangle$
based on the new representation $\quote{\bm{0}}$ also have
$(\sigma_1,\sigma_2,\sigma_3)=(+,+,+)$ and so mix with the
rest of the $(+,+,+)$ sector.  For $N > 3$, we therefore need to
represent the potential using $6{\times}6$ matrices instead of $5{\times}5$
ones.  Table \ref{tab:stNblock} presents the 
corresponding table of $\mtrx{\langle s|t \rangle}$
matrix elements, which can be extracted from the work of
refs.\ \cite{NSZ6j} or \cite{Sjodahl},%
\footnote{
  See our earlier footnote \ref{foot:QCD6j}.  Our table
  \ref{tab:stNblock} corrects for the (non-propagating) misprints
  in ref.\ \cite{NSZ6j} eq.\ (C.17).
  The 6-$j$ symbols involving $\quote{\bm{0}}$ are not explicitly
  presented in the text of
  ref.\ \cite{Sjodahl} but, if using ref.\ \cite{Sjodahl},
  one could infer the $\quote{\bm{0}}$ entries of our table
  \ref{tab:stNblock} from the other entries by the requirement
  that $\mtrx{\langle s|t \rangle}$ be unitary.
  One may also find the 6-$j$ symbols involving $\quote{\bm{0}}$
  in the electronic supplementary materials for ref.\ \cite{Sjodahl}.
  When comparing to either reference, note that our table
  \ref{tab:stNblock} focuses on the $(\sigma_1,\sigma_2,\sigma_3)=(+,+,+)$
  states and so only gives matrix elements involving the combination
  $\tfrac{1}{\sqrt2} \bigl(
     |{\rm ch}_{\bm{10}}\rangle + |{\rm ch}_{\overline{\bm{10}}}\rangle
   \bigr)$
  of the $|{\rm ch}_{\bm{10}}\rangle$ and $|{\rm ch}_{\overline{\bm{10}}}\rangle$
  states instead of for each of those states separately.
}
and
whose computation is also discussed in our appendix \ref{app:6j}.

\begin{table}[t]

\begin{tabular}{|c|ccccc|c|c|}\hline
%\backslashbox{{${\cal R}$}}{{${\cal R}'$}}
$\langle s_{\cal R} | t_{{\cal R}'} \rangle$
  & $\bm{1}$
  & $\quoteaa{\bm{8}}$ & $\quotess{\bm{8}}$
  & $\quote{\bm{10}+\overline{\bm{10}}}$
  & $\quote{\bm{27}}$ & $\quote{\bm{0}}$
  & \shortstack{conversion \\ to $\langle s_{\cal R} | u_{{\cal R}'} \rangle$} \\
\hline
$\bm{1}$
  & $\frac{1}{N^2-1}$
  & $\sqrt{\frac{1}{N^2-1}}$
  & $\sqrt{\frac{1}{N^2-1}}$
  & $\sqrt{\frac{N^2-4}{2(N^2-1)}}$
  & $\frac{N}{2(N+1)} \sqrt{\frac{N+3}{N-1}}$
  & $\frac{N}{2(N-1)} \sqrt{\frac{N-3}{N+1}}$
  & $+$ \\
$\quoteaa{\bm{8}}$
  &
  & $\frac12$
  & $\frac12$
  & $0$
  & $-\frac12 \sqrt{\frac{N+3}{N+1}}$
  & $ \frac12 \sqrt{\frac{N-3}{N-1}}$
  & $-$ \\
$\quotess{\bm{8}}$
  &
  &
  & $\frac{N^2-12}{2(N^2-4)}$
  & $-\sqrt{\frac{2}{N^2-4}}$
  & $\frac{N}{2(N+2)} \sqrt{\frac{N+3}{N+1}}$
  & $-\frac{N}{2(N-2)} \sqrt{\frac{N-3}{N-1}}$
  & $+$ \\
$\quote{\bm{10}+\overline{\bm{10}}}$
  & \multicolumn{3}{c}{ \multirow{2}{*}{(symmetric)} }
  & $\frac12$
  & $- \sqrt{\frac{(N-2)(N+3)}{8(N+1)(N+2)}}$
  & $- \sqrt{\frac{(N+2)(N-3)}{8(N-1)(N-2)}}$
  & $-$ \\
$\quote{\bm{27}}$
  &
  &
  &
  &
  & $\frac{N^2+N+2}{4(N+1)(N+2)}$
  & $\frac{1}{4} \sqrt{\frac{N^2-9}{N^2-1}}$
  & $+$ \\ \hline
$\quote{\bm{0}}$
  & \multicolumn{5}{c|}{(symmetric)}
  & $\frac{N^2-N+2}{4(N-1)(N-2)}$
  & $+$ \\
\hline
\end{tabular}
\caption{
  \label{tab:stNblock}
  The SU($N$) generalization of table \ref{tab:st3block}.
  Since the table is symmetric, we have not bothered to show
  the entries below the main diagonal.
}
\end{table}

In the limit $N{\to}3$, the 4-gluon singlets $|{\rm ch}_{\quote{\bm{0}}}\rangle$
constructed from the $\quote{\bm{0}}$ representation no longer mix
in table \ref{tab:stNblock}
with any of the other singlets and so will completely
decouple from calculations of overlapping formation time effects:
\begin {equation}
   \begin{array}{|c|}
     \hline
     \\[2em]
     \mbox{\hspace{2em}Table \ref{tab:stNblock}\hspace{2em}} \\
     \\[2em] \hline
   \end{array}
   \qquad \xrightarrow{N=3} \qquad
   \begin{array}{|c|c|}
     \hline 
     \mbox{Table \ref{tab:st3block}} &
        \begin{matrix} 0 \\ 0 \\ 0 \\ 0 \\ 0 \end{matrix}
     \\ \hline
     \begin{matrix} \,0\, & \,0\, & \,0\, & \,0\, & \,0\, \end{matrix} &
         1
     \\ \hline
   \end{array}
   ~.
\end {equation}
This is the reason that SU($N$) results for our application
will smoothly approach the SU(3) result as $N{\to}3$.

Note that if
we were interested in SU(2) gauge theory, the situation would be
more complicated.
For SU(2), the decomposition is
$\Adj\otimes\Adj = \bm{3} \otimes \bm{3}
 = \bm{1}_{\rm s} \oplus \bm{3}_{\rm a} \oplus \bm{5}_{\rm s}$,
which in our SU($3$)-based notation (\ref{eq:AdjxAdj}) corresponds to
what we've called
$\bm{1}_{\rm s} \oplus \quotea{\bm{8}} \oplus \quotes{\bm{27}}$.
But the unwanted representations $\quotes{\bm{8}}$,
$\quote{\bm{10}+\overline{\bm{10}}}$ and $\quote{\bm{0}}$ in the SU(2)
case do not
simply decouple if we set $N{=}2$ in table \ref{tab:stNblock}.
Also, some of the entries involving those unwanted representations
are infinite for $N{=}2$.
Since our application of interest is SU(3),
we'll ignore the issue of whether (and how) one can interpolate the
$N{>}3$ results to SU(2).

Similar to SU(3), consistency requires Casimir scaling of
$\hat q$, which gives
\begin {equation}
   \mtrx{\hat q}_{\rm diag}
%   \equiv
%   \left(
%   \begin{array}{ccccc|c}
%      \,0\, &&&&& \\
%      & \hat q_{\Adj} &&&& \\
%      && \hat q_{\Adj} &&& \\
%      &&& \hat q_{\quote{\bm{10}}} && \\
%      &&&& \hat q_{\quote{\bm{27}}} & \\
%      \hline
%      &&&&& \hat q_{\quote{\bm{0}}}
%   \end{array}
%   \right)
   =
   \frac{\hat q_\Adj}{C_\Adj}
   \left(
   \begin{array}{ccccc|c}
      \,0\, &&&&& \\
      & C_{\Adj} &&&& \\
      && C_{\Adj} &&& \\
      &&& C_{\quote{\bm{10}}} && \\
      &&&& C_{\quote{\bm{27}}} & \\
      \hline
      &&&&& C_{\quote{\bm{0}}}
   \end{array}
   \right)
   =
   \hat q_\Adj
   \left(
   \begin{array}{ccccc|c}
      \,0\, &&&&& \\
      & \,1\, &&&& \\
      && \,1\, &&& \\
      &&& \,2\, && \\
      &&&& 2{+}\tfrac{2}{N} & \\
      \hline
      &&&&& 2{-}\tfrac{2}{N}
   \end{array}
   \right)
   .
\end {equation}
The generalizations of the matrices $\mtrx{S}$ and $\mtrx{T}$ 
of (\ref{eq:ST}) are then found to be
\begin {subequations}
\label {eq:STN}
\begin {equation}
   \mtrx{S} \equiv
     \left(
     \begin{array}{ccccc|c}
       ~1 &&&&& \\
       & ~\tfrac12 &&&& \\
       && ~\tfrac12 &&& \\
       &&& ~0 && \\
       &&&& -\tfrac{1}{N} & \\
       \hline
       &&&&& \tfrac{1}{N}
     \end{array}
     \right)
\end {equation}\
and
\begin {equation}
   \mtrx{T} \equiv
     \left(
     \begin{array}{ccccc|c}
       0 & \tfrac{1}{\sqrt{N^2-1}} & 0 & 0 & 0 & 0\\
       \frac{1}{\sqrt{N^2-1}} & 0 & \tfrac14 & 0
           & \tfrac{1}{2N} \sqrt{\tfrac{N+3}{N+1}}
           & \tfrac{1}{2N} \sqrt{\tfrac{N-3}{N-1}} \\
       0 & \tfrac14 & 0 & \tfrac{1}{\sqrt{2(N^2-4)}} & 0 & 0\\
       0 & 0 & \tfrac{1}{\sqrt{2(N^2-4)}} & 0
           & \tau_+
           & \tau_- \\
       0 & \tfrac{1}{2N} \sqrt{\tfrac{N+3}{N+1}} & 0
           & \tau_+ & 0 & 0 \\
       \hline
       0 & \tfrac{1}{2N} \sqrt{\tfrac{N-3}{N-1}} & 0
           & \tau_- & 0 & 0 \\
     \end{array}
     \right) ,
\end {equation}
\end {subequations}
where
\begin {equation}
   \tau_\pm \equiv \tfrac{1}{2N}\sqrt{\tfrac{(N\mp2)(N\pm1)(N\pm3)}{2(N\pm2)}} .
\end {equation}

% ----------------------------------------------------------------------------

\subsection{The large-\boldmath$N$ limit}

In the large-$N$ limit, (\ref{eq:STN}) becomes
\begin {equation}
   \mtrx{S} \to
     \left(
     \begin{array}{cccccc}
       ~1 &&&&& \\
       & \multicolumn{2}{c}{\multirow{2}{*}{$
            \begin{array}{|cc|}
               \hline
               \tfrac12 & \\
               & ~\tfrac12 \\[2pt] \hline
             \end{array}
         $}} &&& \\[4pt]
       &&&&& \\
       &&& ~0 && \\
       &&&& ~0 & \\
       &&&&& ~0
     \end{array}
     \right) ,
   \qquad
   \mtrx{T} \equiv
     \left(
     \begin{array}{cccccc}
       0 & &            & & & \\
         & \multicolumn{2}{c}{\multirow{2}{*}{$
              \begin{array}{|cc|}
                 \hline
                 0 & ~\tfrac14 \\
                 \tfrac14 & ~0 \\[2pt] \hline
               \end{array}
            $}} &&& \\[20pt]
         & &            & 0 & \tfrac1{2\sqrt2} & \tfrac1{2\sqrt2} \\
         & &            & \tfrac1{2\sqrt2} & 0 & 0 \\
       %\hline
         & &            & \tfrac1{2\sqrt2} & 0 & 0
     \end{array}
     \right) ,
\end {equation}
where the boxes highlight the subspace
$\bigl(|s_{\Adj_{\rm aa}}\rangle, |s_{\Adj_{\rm ss}}\rangle\bigr)$.
We see that those two singlets
decouple from all the others.  Since this is the subspace that our
initial 4-gluon state $|s_{\Adj_{\rm aa}}\rangle$ belongs to, we can reduce
the $6{\times}6$ matrix problem to a $2{\times}2$ problem in
the large $N$ limit.
The 4-body potential given by (\ref{eq:V4Casimir}) and (\ref{eq:TdotTST})
is then
\begin {align}
   \mtrx{V}(\b_1,\b_2,\b_3,\b_4) =
      - \frac{i{\hat q}_{\rm A}}8 \Bigl\{
         &
         \begin{pmatrix} 1&0 \\ 0&1 \end {pmatrix}
              \bigl[ (\b_1{-}\b_2)^2 + (\b_3{-}\b_4)^2 \bigr]
\nonumber\\ &
         +
         \tfrac12
         \begin{pmatrix} 1&1 \\ 1&1 \end{pmatrix}
              \bigl[ (\b_1{-}\b_3)^2 + (\b_2{-}\b_4)^2 \bigr]
\nonumber\\ &
         +
         \tfrac12
         \begin{pmatrix} 1&-1 \\ -1&1 \end{pmatrix}
              \bigl[ (\b_1{-}\b_4)^2 + (\b_2{-}\b_3)^2 \bigr]
      \Bigr\} .
\end {align}
This potential can be diagonalized in color-singlet space by switching basis to
\begin {equation}
   |s_{\Adj_\pm}\rangle \equiv
   \tfrac{1}{\sqrt2} \bigl(
      |s_{\Adj_{\rm aa}} \rangle \pm |s_{\Adj_{\rm ss}}\rangle
   \bigr) ,
\end {equation}
with
\begin {subequations}
\label {eq:Vpm}
\begin {align}
   V_+(\b_1,\b_2,\b_3,\b_4) &=
      - \frac{i{\hat q}_{\rm A}}{8} \Bigl[
           (\b_1{-}\b_2)^2 + (\b_2{-}\b_4)^2 + (\b_4{-}\b_3)^2 + (\b_3{-}\b_1)^2
      \Bigr] ,
\label {eq:Vplus}
\\
   V_-(\b_1,\b_2,\b_3,\b_4) &=
      - \frac{i{\hat q}_{\rm A}}{8} \Bigl[
           (\b_1{-}\b_2)^2 + (\b_2{-}\b_3)^2 + (\b_3{-}\b_4)^2 + (\b_4{-}\b_1)^2
      \Bigr] .
\end {align}
\end {subequations}
Eqs.\ (\ref{eq:Vpm}) exactly match the two color routings used in
previous large-$N$ work on overlapping formation times \cite{seq}
for the contribution of diagrams like fig.\ \ref{fig:seqlabels}.

In contrast, previous large-$N$
work \cite{2brem} did {\it not}\/ need to study two different color routings
for diagrams like fig.\ \ref{fig:split2labels}.
In terms of our analysis here, this happens because the 4-particle evolution
in those diagrams starts with $|s_{\Adj_{\rm aa}}\rangle$ and ends
with $\langle t_{\Adj_{\rm aa}}|$.  Using table \ref{tab:stNblock},
the large-$N$ analog of the SU(3) representation (\ref{eq:stvecs})
of these states as vectors is
\begin {equation}
  |s_{\Adj_{\rm aa}}\rangle \to
    \begin{pmatrix} 0 \\ 1 \\ 0 \\ 0 \\ 0 \\ 0 \end{pmatrix}
  \qquad
  \mbox{and}
  \qquad
  |t_{\Adj_{\rm aa}}\rangle \to
    \mtrx{\langle s|t \rangle}
    \begin{pmatrix} 0 \\ 1 \\ 0 \\ 0 \\ 0 \\ 0 \end{pmatrix}
  =
  \begin{pmatrix}
    0 \\ \tfrac12 \\ \tfrac12 \\ 0 \\ -\tfrac12 \\ \tfrac12
  \end{pmatrix} .
\end {equation}
So, ending the evolution with $\langle t_{\Adj_{\rm aa}}|$ in
fig.\ \ref{fig:split2labels} automatically projects out just
the $V_+$ potential (\ref{eq:Vplus}).%
\footnote{
\label{foot:relabelbs}%
  The convention for labeling the four gluons in ref.\ \cite{2brem}
  is slightly different than our convention here of
  fig.\ \ref{fig:split2labels}.  The difference corresponds
  to relabeling
  $(\b_1,\b_2,\b_3,\b_4) \rightarrow (\b_3,\b_2,\b_4,\b_1)$
  in the formula (\ref{eq:Vplus}) for the potential.
}

% ===========================================================================

\section{Relation to Previous Work}
\label{sec:history}

After completing the original version of this paper, we became aware of
earlier, closely related work by Nikolaev, Sch\"afer, and Zakharov
\cite{NSZ6j} and Zakharov \cite{Zakharov6j}.  Here we explain
the similarities and differences of the uses we make of the
6-$j$ coefficients.

The most significant difference is the details of our application.
Nikolaev et al.\ \cite{NSZ6j} were interested in the limit of a
relatively thin QCD medium and the possible color configurations of
(and implications for hadronization of) a $g{\to}gg$ produced pair
of gluons that leave the medium.  Our application, in contrast, is
to corrections, due to overlapping formation times, to {\it double}
splitting $g{\to}ggg$ in a thick medium.  As a result, we need to
ultimately study quantum mechanical harmonic oscillator problems
of the type (\ref{eq:Hform}), laid out in detail in section
\ref{sec:reduction}.

In ref.\ \cite{Zakharov6j}, Zakharov applied some of the formalism
of ref.\ \cite{NSZ6j} to give interesting quantitative estimates of
the rate at which the combined
color state of a pair of gluons produced by single splitting $g{\to}gg$
randomizes as the pair propagates through simple models of an expanding
quark-gluon plasma.  (The same formalism could be used to discuss this
issue in the theorist's case of a static quark-gluon plasma.)
This calculation was done in a ``rigid geometry'' approximation
(sometimes referred to as ``antenna'' approximation) where partons follow
straight-line trajectories and the quantum mechanical evolution of their
transverse wave functions (which are essential for our own
application) are ignored.  For his purposes in that particular
paper, $\b_1{=}\b_3$ and
$\b_2{=}\b_4$ in our conventions here for (\ref{eq:V4Casimir}).%
\footnote{
  Because of different naming conventions, what Zakharov calls the
  $t$-channel in ref.\ \cite{Zakharov6j} we call the $u$-channel.
}

Another difference of our work is that refs.\ \cite{NSZ6j,Zakharov6j}
treat the medium in a way that is based on a weakly-coupled picture
of the medium.  They treat the medium as a set of static scattering
centers, and the cross-section for scattering from those scattering
centers is determined by single gluon exchange.%
\footnote{
  The fact that they use {\it static} scattering centers is inessential.
  If desired, weak-coupling analysis with static scattering centers
  can be easily generalized to dynamic scattering centers by replacing
  $n\,d\sigma$ by a weak-coupling calculation of the differential
  rate $d\Gamma_{\rm el}$ for
  elastic scattering.  See section II.B of ref.\ \cite{simple}.
}
In contrast, we have focused on the $\hat q$ approximation, which
can (with caveats) be applied to strongly-coupled as well as weakly-coupled
plasmas.  For the weakly-coupled case, ref.\ \cite{NSZ6j} finds
a 4-gluon potential that is equivalent to our (\ref{eq:V4Casimir}) and
(\ref{eq:TdotTST}) but with our $-\frac{i}{4} \hat q_\Adj (\b_i{-}\b_j)^2$
replaced
by the full 2-gluon potential, which in their language corresponds to
replacing our $\tfrac14 \hat q_\Adj (\b_i{-}\b_j)^2$ by $n\,\sigma(\b_i{-}\b_j)$.
For weak coupling, our version of the potential just represents
making the usual $\hat q$ quadratic approximation to their
$\sigma(b)$ for small $b$.  In that case, Casimir scaling of
$\hat q$ or of $\sigma(b)$
with the color representation of the high-energy particle
is automatic, following easily from weak-coupling calculations
of those quantities.
We wanted our approach to also apply to strongly-coupled
plasmas, and so we did not assume that $\hat q$ obeyed Casimir scaling.
Instead,
one of the results we found by constructing the 4-gluon potential
was that self-consistency of the $\hat q$ approximation necessarily
requires such scaling.  We have avoided going beyond the $\hat q$
approximation simply because we do not have any arguments for how
that would work for a strongly-coupled medium.
In particular, the 4-gluon potential need not then be exactly expressible
in terms of 2-gluon potentials.

Ref.\ \cite{NSZ6j} noted that all but five color states decouple
from the problem; we believe that our symmetry classification
$(\sigma_1,\sigma_2,\sigma_3)$ of states under channel-preserving
permutations (as in Table \ref{tab:signs})
provides a useful way to understand this decoupling.

% ===========================================================================

\section{Conclusion}
\label{sec:conclusion}

Calculations in the literature of the effect of overlapping formation times
on in-medium shower development have resorted to either soft bremsstrahlung
or large-$N$ limits.  Previous large-$N$ calculations, that avoid
soft-bremsstrahlung approximations, use propagators for 4-particle evolution
in figs.\ \ref{fig:split2labels} and \ref{fig:seqlabels}
mathematically equivalent to the propagator for
a system of two non-relativistic (and non-Hermitian)
coupled harmonic oscillators.
We've shown here that the large-$N$ limit can be avoided
at the cost of replacing the ``spring constants'' of that
harmonic oscillator problem by
constant $5{\times}5$ matrices, which have been explicitly derived in this
paper for $g \to ggg$ processes.
These constant matrices act on an internal color space of 4-gluon
color singlets.  Unfortunately, the two matrices $\mtrx{S}$ and $\mtrx{T}$
(\ref{eq:ST}) used in their construction (\ref{eq:HVeffective})
do not commute outside of
the large-$N$ limit, and so the harmonic oscillator problem
(\ref{eq:Hform2}) cannot be solved by straightforward diagonalization
in color-singlet space.
We do not know how to find a closed-form solution for
this ``harmonic oscillator'' propagator.
One could, of course, instead solve for the propagator numerically by
solving the corresponding Schr\"odinger equation numerically, which
would be
\begin {equation}
  i \partial_t \mtrx{G}(q_1,q_2,t;q'_1,q'_2,0)
  =
  \left[
    - \frac{\partial_{q_{1}}^2}{2m_1}
    - \frac{\partial_{q_{2}}^2}{2m_2}
    + \frac12
      \begin{pmatrix} q_{1} \\ q_{2} \end{pmatrix}^{\!\!\top}
      \begin{pmatrix} \mtrx{a} & \mtrx{b} \\ \mtrx{b} & \mtrx{c} \end{pmatrix}
      \begin{pmatrix} q_{1} \\ q_{2} \end{pmatrix}
  \right]
  \mtrx{G}(q_1,q_2,t;q'_1,q'_2,0)
\label {eq:Gschro}
\end {equation}
[where $q_1$ and $q_2$ are two-dimensional vectors as made explicit
in (\ref{eq:Hform2})].
Initial conditions would be set by
either (i) appropriately normalized delta functions
$\delta(q_1{-}q'_1)\,\delta(q_2{-}q'_2)$
or, perhaps more usefully, (ii) the functions the propagator is
ultimately convolved
with in the application \cite{2brem,seq} (see below).

However, use of a numerical solution for the propagator will be complicated.
For the large-$N$ case, it was possible to use the known analytic form
of a standard harmonic oscillator propagator to analytically perform most of
the integrals in calculations of the overlap effects.
Consider, for example, the overlap effects on the differential rate
$d\Gamma/dx\,dy$ for $g \to ggg$ with daughter energy fractions
$x$, $y$, and $1{-}x{-}y$.
The calculation of that rate involves integrals of the form,
for example,%
\footnote{
  The specific example (\ref{eq:nasty}) corresponds to
  fig.\ \ref{fig:seqlabels} and is taken from eq.\ (E7) of ref.\ \cite{seq},
  with various variables relabeled here, including the particle labels
  1234 as described in footnote \ref{foot:relabelbs}.
  For a slightly different but closely related example
  corresponding to fig.\ \ref{fig:split2labels},
  see eq.\ (5.10) of ref.\ \cite{2brem}.
}
\begin {multline}
   \int_0^\infty d(\Delta t)
   \int_{\B',\B}
   \frac{B'_{\bar n}}{(B')^2} \,
   \frac{B_m}{B^2} \,
   \exp\bigl(
      - \tfrac12 |M'| \Omega' (B')^2
      - \tfrac12 |M| \Omega (B)^2
   \bigr)
\\ \times
   \nabla_{\bcalC'_{34}}^{\bar m}
   \nabla_{\bcalC_{12}}^n
   G(\bcalC'_{12},\bcalC'_{34},\Delta t;\bcalC_{12},\bcalC_{34})
   \Bigr|_{\bcalC'_{34}=0=\bcalC_{12}; ~ \bcalC_{12}'=\B'; ~ \bcalC_{34}=\B} .
\label {eq:nasty}
\end {multline}
With a standard harmonic oscillator propagator, it was possible to do
the two 2-dimensional integrals over $\B$ and $\B'$ analytically, leaving
only the single $\Delta t$ integral for numerical integration.
But with only a numerical result for the propagator, one would have to
do the $\B$ and $\B'$ integrals numerically as well.
Moreover, to make use of the rates to calculate the development of
showers, one also needs $x$ and $y$ integrals
involving $d\Gamma/dx\,dy$ (see ref.\ \cite{qedNfstop}, for example).
So, if the propagator is found numerically using (\ref{eq:Gschro}),
that numerical result would ultimately need to be used inside a
numeric integration over $(\B,\B',\Delta t,x,y)$.
That's not impossible but, we think, likely complicated to do accurately.

It would be {\it very} useful if there were, after all, some
way to find an analytic solution to (\ref{eq:Gschro}) for the propagator.
Or alternatively, perhaps, to find some efficient expansion
of the solution for which some integrals in (\ref{eq:nasty}),
such as $(\B,\B')$,
could be done analytically term by term.

% =========================================================================
% =========================================================================

\begin{acknowledgments}

My profound thanks to Han-Chih Chang who participated in the early stages
of this project.
He demurred when recently offered co-authorship upon completion
of this paper
but is the reason this paper was written with the stylistic
choice of using first-person plural.

I am also grateful to Howard Haber for discussions about
SU($N$) group theory, and to
Diana Vaman, Dima Pesin, Malin Sjodahl, Nikolai Nikolaev, and Bronislav
Zakharaov for other useful conversations.
This work was supported, in part, by the U.S. Department
of Energy under Grant No.~DE-SC0007984.

\end{acknowledgments}

% =========================================================================
\appendix
% =========================================================================

\section{6-\boldmath$j$ coefficients for four gluons in SU(\boldmath$N$)}
\label {app:6j}

In this appendix, we give explicit constructions for the
$s$-channel basis of
4-gluon singlet states, with primary focus on the five
$(\sigma_1,\sigma_2,\sigma_3){=}(+,+,+)$ singlets relevant to our application.
Then we outline our calculation of the relevant 6-$j$ coefficients.

% --------------------------------------------------------------------------

\subsection{Projections that decompose \boldmath$\Adj\otimes\Adj$}

We start by discussing how to combine two gluons into a color
representation $R$.
The singlet state ($\bm{1}_{\rm s}$)
of two gluons is proportional to $\delta_{AB} |A B\rangle$.
One way to make an adjoint state ($\Adj_{\rm a}$)
is proportional to $f_{ABc} |AB\rangle$, where
$f_{abc}$ are the Lie algebra structure constants, given in terms of
the fundamental-representation generators $T^a_{\rm F}$ as
$f_{abc} = \tfrac{2}{i} \tr([T^a_{\rm F},T^b_{\rm F}]T^c_{\rm F})$.
Here and throughout, we use the typical particle physics (as opposed to
mathematics) normalization convention for the generators that
$\tr(T^a_{\rm F} T^b_{\rm F}) = \frac12 \delta^{ab}$.
Another way to make an adjoint state ($\Adj_{\rm s}$) is
$d_{ABc} |AB\rangle$, in terms of the completely symmetric
$d_{abc} \equiv 2 \tr(\{T^a_{\rm F},T^b_{\rm F}\}T^c_{\rm F})$.

We'll find it convenient to write down the corresponding projection
operators ${\cal P}_R$ on the space $\Adj\otimes\Adj$, such that
\begin {equation}
   |a b\rangle =
   (
     {\cal P}_{\bm{1}} +
     {\cal P}_{\Adj_{\rm a}} +
     {\cal P}_{\Adj_{\rm s}} +
     {\cal P}_{\quote{\bm{10}}} +
     {\cal P}_{\quote{\overline{\bm{10}}}} +
     {\cal P}_{\quote{\bm{27}}} +
     {\cal P}_{\quote{\bm{0}}}
  )_{ab,AB} |A B\rangle .
\end {equation}
The first three are
\begin {subequations}
\label {eq:P1AA}
\begin {align}
  ({\cal P}_{\bm 1})_{ab,AB} &= \tfrac{1}{\dA} \delta_{ab} \delta_{AB} \,, \\
  ({\cal P}_{\Adj_{\rm a}})_{ab,AB} &= \tfrac{1}{\CA} f_{abc} f_{ABc} \,, \\
  ({\cal P}_{\Adj_{\rm s}})_{ab,AB} &= \tfrac{1}{\Delta} d_{abc} d_{ABc} \,,
\end {align}
\end {subequations}
where $\dA$ and $\CA$ are the dimension and quadratic Casimir
of the adjoint representation, and we've defined $\Delta$ by
\begin {equation}
   d_{abc} d_{abd} = \Delta \, \delta_{cd} \,,
\end {equation}
which is analogous to the relation
\begin {equation}
   f_{abc} f_{abd} = \CA \delta_{cd} \,.
\end {equation}
For SU($N$),
\begin {equation}
  \dA={N^2-1}, \qquad \CA=N, \qquad \Delta = \frac{N^2-4}{N} \,,
\end {equation}
As appropriate to projection operators, (\ref{eq:P1AA}) is normalized
so that each $({\cal P}_R)^2 = {\cal P}_R$.

To find projection operators for the other states, split
$\Adj\otimes\Adj$ into its symmetric and anti-symmetric combinations:
\begin {align}
   (\Adj\otimes\Adj)_{\rm s} &=
   \bm{1} \oplus \Adj_{\rm s} \oplus \quote{\bm{27}} \oplus \quote{\bm{0}} ,
\\
   (\Adj\otimes\Adj)_{\rm a} &=
   \Adj_{\rm a} \oplus \quote{\bm{10}} \oplus \quote{\overline{\bm{10}}} .
\end {align}
We may then project $\quote{\bm{10} \oplus \overline{\bm{10}}}$
by isolating the subspace of anti-symmetric color
states and subtracting the part that belongs to $\Adj_{\rm a}$:
\begin {align}
  ({\cal P}_{\quote{\bm{10}\oplus\overline{\bm{10}}}})_{ab,AB} \equiv
  ({\cal P}_{\quote{\bm{10}}} + {\cal P}_{\quote{\overline{\bm{10}}}})_{ab,AB} &=
  \tfrac12 (\delta_{aA} \delta_{bB} - \delta_{aB} \delta_{aA})
  - ({\cal P}_{\Adj_{\rm a}})_{ab,AB}
\nonumber\\
  &=
  \tfrac12 (\delta_{aA} \delta_{bB} - \delta_{aB} \delta_{aA})
  - \tfrac{1}{\CA} f_{abc} f_{ABc}
  \,.
\label {eq:Ppsipsibar}
\end {align}
We'll find later that we need not pick out
$\quote{\bm{10}}$ and $\quote{\overline{\bm{10}}}$ separately,
and so (\ref{eq:Ppsipsibar}) will be good enough.%
\footnote{
  If desired, the $\quote{\bm{10}}$ and
  $\quote{\overline{\bm{10}}}$ could be separated using the methods
  of this appendix
  by following steps analogous to (\ref{eq:Zdef}--\ref{eq:P27}).
  If one finds an SU($N$)-covariant operator ${\cal Z}$ that acts
  non-trivially on the subspace $\quote{\bm{10}\oplus\overline{\bm{10}}}$
  with the property that ${\cal Z}^2 = -1$ on that subspace, then
  $\frac12 (1\mp i{\cal Z}) {\cal P}_{\quote{\bm{10}\oplus\overline{\bm{10}}}}$
  gives two projection operators,
  dividing $\quote{\bm{10}\oplus\overline{\bm{10}}}$ into
  the two conjugate complex representations.
  One can show that the operator
  ${\cal Z}_{ab,AB} \equiv \tfrac12 (f_{bAc} d_{aBc} + d_{bAc} f_{aBc})$
  has the desired property.
}

Similarly, we can project $\quote{\bm{27}\oplus\bm{0}}$ by
isolating the subspace of symmetric states and then subtracting
the pieces corresponding to the singlet and $\Adj_{\rm s}$:
\begin {align}
  ({\cal P}_{\quote{\bm{27}\oplus\bm{0}}})_{ab,AB} &=
  \tfrac12 (\delta_{aA} \delta_{bB} + \delta_{aB} \delta_{aA})
  - ({\cal P}_{\bm{1}})_{ab,AB}
  - ({\cal P}_{\Adj_{\rm s}})_{ab,AB}
\nonumber\\
  &=
  \tfrac12 (\delta_{aA} \delta_{bB} + \delta_{aB} \delta_{aA})
  - \tfrac{1}{\dA} \delta_{ab} \delta_{AB}
  - \tfrac{1}{\Delta} d_{abc} d_{ABc}
  \,.
\label {eq:P270}
\end {align}
We need to isolate the two different representations
$\quote{\bm{27}}$ and $\quote{\bm{0}}$.
If we find an SU($N$)-covariant
operator $Z$ that acts non-trivially on the subspace
$\quote{\bm{27}\oplus\bm{0}}$ with the property that $Z^2{=}1$ on that
subspace, then $\frac12 (1 \mp Z) {\cal P}_{\quote{\bm{27}\oplus\bm{0}}}$
would give two projection operators that split the subspace
into $\quote{\bm{27}}$ and $\quote{\bm{0}}$.
One may verify that the operator
\begin {equation}
   Z_{ab,AB} \equiv
   \tfrac12 (f_{aAc} f_{bBc} + f_{aBc} f_{bAc})
\label {eq:Zdef}
\end {equation}
has square
\begin {equation}
   (Z^2)_{ab,AB}
   = Z_{ab,\alpha\beta} Z_{\alpha\beta,AB}
   = \tfrac12 (\delta_{aA}\delta_{bB} + \delta_{aB}\delta_{bA})
     + \tfrac{N}{4} d_{abc} d_{ABc} + \delta_{ab}\delta_{AB}
\end {equation}
for $SU(N)$, so that $Z^2$ acts as the identity on
$\quote{\bm{27}\oplus\bm{0}}$.
The resulting projection operators are
\begin {align}
  ( {\cal P}_{\substack{ \quote{\bm{27}} \\ \quote{\bm{0}} }} )_{ab,AB}
   &=
   \tfrac14 (\delta_{aA} \delta_{bB} + \delta_{aB} \delta_{aA})
   \mp \tfrac14 (f_{aAc} f_{bBc} + f_{aBc} f_{bAc})
\nonumber\\ & \qquad
   + (-\tfrac12 \pm \tfrac{N}{2}) ({\cal P}_{\bm 1})_{ab,AB}
   + (-\tfrac12 \pm \tfrac{N}{4}) ({\cal P}_{\Adj_{\rm s}})_{ab,AB} \,,
\label {eq:P27}
\end {align}
where the upper signs refer to $\quote{\bm{27}}$ and the lower to
$\quote{\bm{0}}$.

% --------------------------------------------------------------------------

\subsection{The 4-gluon color singlets}

The $(\sigma_1,\sigma_2,\sigma_3){=}(+,+,+)$ color singlets have a simple
construction in terms of the projection operators: they are just
\begin {equation}
  |s_{\cal R}\rangle \propto ({\cal P}_{\cal R})_{ABCD} |ABCD\rangle ,
\label{eq:schannel}
\end {equation}
with the clarification that one should identify
\begin {align}
  |s_{\Adj_{\rm aa}}\rangle &\propto ({\cal P}_{\Adj_{\rm a}})_{AB,CD} |ABCD\rangle ,
\\
  |s_{\Adj_{\rm ss}}\rangle &\propto ({\cal P}_{\Adj_{\rm s}})_{AB,CD} |ABCD\rangle ,
\\
  |s_{\quote{\bm{10}+\overline{\bm{10}}}}\rangle
  &\equiv
  \tfrac{1}{\sqrt2} \Bigl(
     |s_{\quote{\bm{10}}}\rangle + |s_{\quote{\overline{\bm{10}}}}\rangle
    \Bigr)
    \propto ({\cal P}_{\quote{\bm{10}\oplus\overline{\bm{10}}}})_{AB,CD} |ABCD\rangle .
\end {align}
The overlap matrix elements of table \ref{tab:stNblock} are then
\begin {equation}
  \langle s_{\cal R} | t_{\cal R'} \rangle
  =
  \frac{ (P_{\cal R})_{AB,CD} (P_{\cal R'})_{AC,BD} }
       { [ (P_{\cal R})_{ab,cd} (P_{\cal R})_{ab,cd}) ]^{1/2}
         [ (P_{\cal R'})_{\alpha\beta,\gamma\delta}
           (P_{\cal R'})_{\alpha\beta,\gamma\delta}) ]^{1/2} }
  \,,
\label {eq:stoverlap}
\end {equation}
where the denominator accounts for normalization of the
4-gluon states (\ref{eq:schannel}).  They may be computed using
the following SU($N$) formulas for the contraction of four or fewer
$f_{abc}$'s and $d_{abc}$'s.

% --------------------------------------------------------------------------

\subsection{Useful SU(\boldmath$N$) relations}

Useful SU($N$) relations for the preceding calculations include
\cite{Azcarraga}
\begin {align}
  \tr(D^a) &= 0 ,
\\
  \tr(F^a F^b F^c F^d) &=
    \delta_{ab} \delta_{cd} + \delta_{ad} \delta_{bc}
    + \tfrac{N}{4} (d_{abe} d_{cde} - d_{ace} d_{bde} + d_{ade} d_{bce}) ,
\\
  \tr(F^a F^b D^c D^d) &=
    \tfrac{(N^2-4)}{N^2} (\delta_{ab} \delta_{cd} - \delta_{ac} \delta_{bd})
    + \tfrac{(N^2-8)}{4N} (d_{abe} d_{cde} - d_{ace} d_{bde})
    + \tfrac{N}{4} d_{ade} d_{bce} ,
\\
  \tr(F^a D^b F^c D^d) &=
    \tfrac{N}{4} (d_{abe} d_{cde} - d_{ace} d_{bde} + d_{ade} d_{bce}) ,
\\
  \tr(D^a D^b D^c D^d) &=
    \tfrac{(N^2-4)}{N^2} (\delta_{ab} \delta_{cd} + \delta_{ad} \delta_{bc})
    + \tfrac{(N^2-16)}{4N} (d_{abe} d_{cde} + d_{ade} d_{bce})
    - \tfrac{N}{4} d_{ace} d_{bde} ,
\\
  [D^a,D^b]_{cd} &=
    i f_{abe} (F^e)_{cd}
    - \tfrac{2}{N} (\delta_{ac} \delta_{bd} - \delta_{ad} \delta_{bc}) ,
\end {align}
where
\begin {equation}
  (F^a)_{bc} \equiv -i f_{abc} ,
  \qquad
  (D^a)_{bc} \equiv d_{abc} .
\end {equation}
Ref.\ \cite{Haber} gives a very useful summary of these and many
other relations, drawn from (and correcting a few typographic errors in)
refs.\ \cite{Kaplan,Azcarraga,MacFarlane,Fadin,Cutler,Tarjanne}.

%%%%%%%%%%%%%%%%%%%%%%%%%%%%%%%%%%%%%%%%%%%%%%%%%%%%%%%%%%%%%%%%%%%%%%%%%%%%%%

\end {document}